\documentclass[lettersize,journal]{IEEEtran}
\usepackage{cite}
\usepackage{graphicx}
\graphicspath{{figures/}}
\usepackage[justification=centering]{caption}
\usepackage[caption=false,font=footnotesize,labelfont=rm,textfont=rm]{subfig}
\usepackage{float}
\usepackage{subfig}
\usepackage{setspace}
\usepackage{bm}
\usepackage{xcolor}
\usepackage{amsmath}
\usepackage{amssymb}
\usepackage{amsfonts}
\usepackage{algorithm}
\usepackage{algpseudocode}
\usepackage{mathtools}
\newtheorem{remark}{\it Remark}[section]

\newtheorem{proposition}{\it Proposition}[section]
\newtheorem{lemma}{\it Lemma}[section]

\definecolor{red}{RGB}{153,0,103}
\definecolor{blue}{RGB}{103,0,153}

\def\BibTeX{{\rm B\kern-.05em{\sc i\kern-.025em b}\kern-.08em
    T\kern-.1667em\lower.7ex\hbox{E}\kern-.125emX}}

\begin{document}
\title{Adaptive Source-Channel Coding for Multi-User Semantic and Data Communications}
\author{Kai~Yuan, Dongxu~Li, Jianhao~Huang, Han~Zhang, and Chuan~Huang
	\thanks{
	{K. Yuan and D. Li are with the Shenzhen Future Network of Intelligence Institute, the School of Science and Engineering, and the Guangdong Provincial Key Laboratory of Future Networks of Intelligence, The Chinese University of Hong Kong, Shenzhen 518172, China (e-mail: kaiyuan3@link.cuhk.edu.cn and dongxuli@link.cuhk.edu.cn). }
	
	{J. Huang is with the Department of Electrical and Electronic Engineering, the University of Hong Kong, Hong Kong 999077, China (e-mail: jianhaoh@hku.hk).}
	
	{H. Zhang is with the Key Laboratory of Universal Wireless Communications, Ministry of Education, Beijing University of Posts and Telecommunications, Beijing 100876, China (e-mail: hanzh92@bupt.edu.cn).}
	
	{C. Huang is with the School of Science and Engineering, the Shenzhen Future Network of Intelligence Institute, and the Guangdong Provincial Key Laboratory of Future Networks of Intelligence, The Chinese University of Hong Kong, Shenzhen 518172, China (e-mail: huangchuan@cuhk.edu.cn).}	
	}
}
\maketitle
\begin{abstract}
This paper considers a multi-user semantic and data communication (MU-SemDaCom) system, where a base station (BS) simultaneously serves users with different semantic and data tasks through a downlink multi-user multiple-input single-output (MU-MISO) channel. The coexistence of heterogeneous communication tasks, diverse channel conditions, and the requirements for digital compatibility poses significant challenges to the efficient design of MU-SemDaCom systems. To address these issues, we propose a multi-user adaptive source-channel coding (MU-ASCC) framework that adaptively optimizes deep neural network (DNN)-based source coding, digital channel coding, and superposition broadcasting according to the channel conditions. First, we employ a data-regression method to approximate the end-to-end (E2E)  semantic and data distortions, for which no closed-form expressions exist due to the complex coupling between DNN-based source coding and channel codes. The obtained logistic formulas decompose the E2E distortion as the addition of the source and channel distortion terms, in which the logistic parameter variations are task-dependent and jointly determined by both the DNN and channel parameters. Then, based on the derived formulas, we formulate a weighted-sum E2E distortion minimization problem that jointly optimizes the source-channel coding rates, power allocation, and beamforming vectors for both the data and semantic users.
Finally, an alternating optimization (AO) framework is developed, where the adaptive rate optimization is solved using the subgradient descent method, while the joint power and beamforming is addressed via the uplink-downlink duality (UDD) technique. Simulation results demonstrate that, compared with the conventional separate source-channel coding (SSCC) and deep joint source-channel coding (DJSCC) schemes that are designed for a single task, the proposed MU-ASCC scheme achieves simultaneous improvements in both the data recovery and semantic task performance.
\end{abstract}
 \vspace{-0.5cm} 

\section{Introduction}
The rapid proliferation of multimedia applications in the sixth-generation (6G) wireless networks, such as augmented/extended reality (AR/XR), autonomous vehicles, and remote operations, poses tremendous challenges to conventional communication systems,  particularly in sustaining high efficiency under limited spectrum resources \cite{yang2022semantic}. To tackle these challenges, semantic communication (SemCom) offers a paradigm-shifting solution that aims to transmit the underlying meaning of source data rather than delivering raw bits, thereby significantly reducing communication overheads \cite{joda2022internet}.
Unlike conventional system design, which emphasizes accurate bit-level transmission, SemCom integrates source and channel coding to directly minimize end-to-end (E2E) distortion \cite{gunduz2024joint}.
However, in a multi-user system, SemCom needs to simultaneously serve users with diverse communication tasks while maintaining compatibility with modern communication hardware, presenting new challenges for efficient SemCom system design \cite{guo2024survey,lin2023pushing}.

One of the typical techniques in SemCom is the deep joint source-channel coding (DJSCC) approach, which employs the deep neural networks (DNNs) to extract and transmit low-dimensional features of source data in an E2E framework.
While early DJSCC works \cite{xie2021deep,bourtsoulatze2019deep,dai2022nonlinear,10529950} achieved superior E2E performance over conventional separate source-channel coding (SSCC) schemes, they rely on analog signal transmission, which is fundamentally incompatible with modern digital wireless systems. 
To address this issue, digital SemCom systems have been developed to improve compatibility with practical systems. 
The authors in \cite{tung2022deepjscc,bo2024joint,hu2023robust} introduced DNN-based quantization modules into the analog DJSCC architecture, which maps semantic features into digital representations while maintaining the differentiability during E2E training. 
However, these approaches require manually designing the quantization strategies and cannot adapt to the variations of sources or channels. To better leverage the power of digital codes, recent studies have investigated the weakly-coupled JSCC design, where the source and channel codings are separately designed but jointly optimized for E2E distortion reduction \cite{huang2025d,li2025adaptive}. Specifically, the authors in \cite{huang2025d} proposed a digital deep source-channel coding architecture, where the deep neural network (DNN) parameters and the digital channel coding rate are jointly optimized to minimize the mean squared error (MSE) in image transmission. Building on this line of research, the authors in \cite{li2025adaptive} proposed an adaptive source-channel coding (ASCC) framework that jointly optimizes source and channel rates to achieve channel adaptability and minimize semantic distortion.

Inspired by the success of single-user SemCom systems, recent studies have begun exploring multi-user SemCom (MU-SemCom) systems by employing multiple access (MA) techniques \cite{clerckx2024multiple}. 
Orthogonal multiple access (OMA) schemes, including orthogonal frequency division multiple access (OFDMA), time division multiple access (TDMA), and orthogonal space division multiple access (OSDMA) \cite{rom2012multiple}, have been utilized in MU-SemCom systems to achieve interference mitigation by exclusively assigning distinct time slots, frequency bands, or spatial dimensions to different users.
In particular, the authors in \cite{zhang2023optimization}  considered the OFDMA system and developed a reinforcement learning framework to design the resource block allocation policy that maximizes the image-to-graph semantic similarity.
Similarly, the TDMA was adopted in \cite{kang2022personalized} to transmit the semantic triplets to receivers and a power allocation module was introduced based on the personalized priorities of the triplets.
Moreover, based on the encoder-decoder architecture as in \cite{bourtsoulatze2019deep}, the OSDMA technique was employed to convert the multi-user interference channel into a parallel channel through zero-forcing (ZF) beamforming without considering the interferences \cite{xu2023semantic,xie2021task,zhang2022multi}. 
However, these OMA-based schemes inherently limit spectral efficiency due to their exclusive resource allocation to a single user, lacking the capability to adaptively share time, frequency, or spatial resources based on specific task requirements and channel conditions \cite{dai2018survey}.

To address this limitation, recent research efforts have investigated non-orthogonal multiple access (NOMA)-based MU-SemCom systems that enable concurrent transmissions via shared time, frequency, and spatial resources \cite{mu2022heterogeneous,zhang2024non,mu2023exploiting,zhang2023model,liang2024orthogonal,zhang2025beamforming,zhao2024joint}.
The authors in \cite{mu2022heterogeneous,zhang2024non,mu2023exploiting} have developed NOMA-powered two-user SemCom systems leveraging the successive interference cancellation (SIC) decoding.
Additionally, research advances in model division multiple access (MDMA) \cite{zhang2023model,liang2024orthogonal} demonstrated innovative utilization of semantic information subspaces, implementing interference mitigation mechanisms through intra-model and inter-model orthogonal projections to achieve significant bandwidth efficiency improvements.
The authors in \cite{zhang2025beamforming} proposed a beamforming design method in a semantic and bit user coexisting system and outperformed the conventional ZF, maximum ratio transmission (MRT), and weighted minimum mean-square error (WMMSE) methods. 
Despite these advancements, existing MU-SemCom systems still face challenges when integrated with digital hardware, as they encounter deployment limitations due to their reliance on analog signal transmission.
Although the works in \cite{zhang2023optimization} and \cite{zhao2024joint} considered digital source-channel coding methods, they adopted the Shannon capacity as channel coding rate, which is unattainable in practical finite blocklength channel coding regimes. 
The authors in \cite{gao2024semantic} proposed an adaptive channel coding rate method in the multi-user modality fusion task and employed the finite blocklength channel coding. 
However, this method used fixed source coding modules and focused on the modality fusion task, which restricts system efficiency by lacking source-channel adaptation to diverse task requirements and channel conditions. 

This paper aims to propose a multi-user ASCC (MU-ASCC) framework for the digital multi-user semantic and data communication (MU-SemDaCom) system to simultaneously serve users with different data and semantic tasks over the same frequency band. 
In particular, we consider a downlink multi-user multiple-input single-output (MU-MISO) system where the multi-antenna base station (BS) employs DNNs to extract user-specific semantic features, followed by digital source-channel coding and superposition coding \cite{cover1972broadcast} for simultaneous transmissions to single-antenna users. 
Unlike most of the aforementioned MU-SemCom works considering homogeneous users with identical tasks \cite{zhao2024joint, gao2024semantic}, we consider a MU-SemDaCom scenario, where the served users can be divided into two categories: data users (DUs) aiming for source data reconstruction and semantic users (SUs) for semantic task execution \cite{zhang2022multi}. 
The MU-ASCC adaptively optimizes source-channel coding rates together with resource allocation, aiming to minimize the overall E2E distortions of both the SUs and DUs in MU-SemDaCom systems.

The key contributions and findings of this paper are summarized as follows:
\vspace{-0.05cm} 
\begin{enumerate}
	\item \textbf{E2E Distortions of MU-SemDaCom}: 
	To facilitate the E2E performance analysis, we establish the analytical models of the E2E distortions for users with different tasks in the MU-SemDaCom system. Unlike the single-user system \cite{li2025adaptive},  the E2E distortions of the MU-SemDaCom system depend on both the channel noise and inter-user interference. First, we approximate the bit error rate (BER) as a function of signal-to-interference-plus-noise ratio (SINR) and channel coding rate based on the finite blocklength transmission theory \cite{polyanskiy2010channel}. Then, we approximate the E2E distortions for both the DUs and SUs as logistic functions of BER and source coding rate according to the empirical results over widely-studied datasets.
	The E2E distortion formulas reveal that different tasks require different source-channel coding rates and exhibit varying levels of tolerance to BER. This inherent task diversity provides the foundation for adaptive optimization in the heterogeneous MU-SemDaCom system.
	\item \textbf{Joint Rate, Power and Beamforming Optimization}: Based on the E2E distortions,
	we formulate a joint optimization problem to adaptively optimize the source and channel coding rates, transmission power, and beamforming according to channel conditions and task-specific characteristics, with the objective of minimizing the weighted-sum E2E distortion under the power budget and transmission delay constraints. To solve this problem, we develop an alternating optimization (AO) algorithm to decompose the joint optimization into two subproblems: adaptive source-channel rate optimization and joint power and beamforming optimization. The adaptive source-channel optimization allocation problem is reformulated as multiple parallel single-variable optimizations. For the joint power and beamforming optimization, we leverage the uplink-downlink duality (UDD) theory to transform it into an equivalent uplink problem, which can be efficiently solved by the AO algorithm. 
	\item \textbf{Experiments}: Experimental results reveal that the proposed method outperforms both the traditional SSCC scheme and DJSCC scheme. Specifically, our approach simultaneously enhances data reconstruction performance (measured in multi-scale structural similarity index (MS-SSIM)) for DUs and semantic task execution performance (measured in classification accuracy) for SUs through adaptive optimization of the source-channel coding, power allocation and beamforming in response to channel conditions. Furthermore, our framework characterizes the achievable performance region of the MU-SemDaCom system by adjusting the distortion weight of each user. Within this region, both the DUs and SUs can simultaneously achieve superior performances compared to the benchmarking schemes. This performance gain stems from the powerful feature extraction capability of the DNN-based source coding and the adaptive optimization of resource allocation according to channel conditions and task-specific characteristics.
\end{enumerate}

The remainder of this paper is organized as follows. Section \ref{section_system_model} introduces the MU-SemDaCom system model. Section \ref{section_problem_formulation} characterizes the E2E distortion and formulates the optimization problem. The proposed joint rate, power and beamforming (JRPB) optimization algorithm is proposed in Section \ref{section_joint_rate_power_beam}. Simulation results are shown in Section \ref{section_experiment} and Section \ref{section_conclusion} concludes this article.

Notations: Lowercase and uppercase letters, e.g., $x$ and $M$, denote scalars; Boldface letters, e.g., $\bm{x}$, denote vectors; $\lceil \cdot \rceil$ denotes the celling operation; $||\bm{x}||$ represents the 2-norm of vector $\bm{x}$ and $|\cdot|$ represents the norm of a complex number; $\log(\cdot)$ and $\log_n(\cdot)$ are the logarithm functions with base $e$ and $n$, respectively; $\bm{1}_n$ is a $n$-length column vector with all elements being 1; $\bm{I}_n$ is the identity matrix with dimension $n\times n$; $\mathbb{R}^n$ and $\mathbb{C}^n$ are the real and complex vector space with dimension $n$, respectively; $\mathbb{E}_{x}\{\cdot\}$ denotes the expectation operation with respect to $x$.
\vspace{-0.3cm} 
\section{System Model}
\label{section_system_model}
In this section, we present the considered MU-SemDaCom system, followed by E2E distortion evaluations.
\vspace{-0.3cm} 
\subsection{MU-SemDaCom System}
\begin{figure}[ht]
	\centering
	\includegraphics[width=0.7\linewidth]{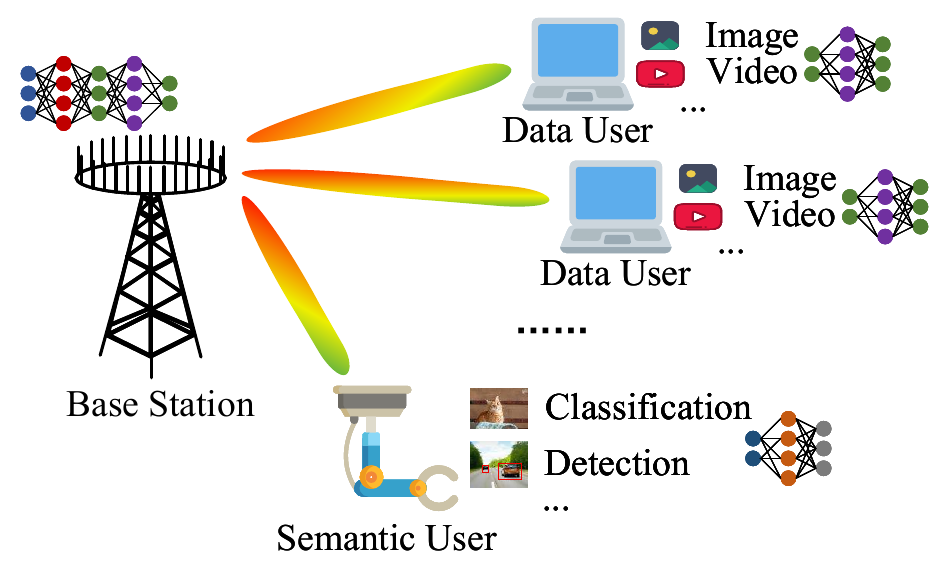}
	\vspace{-0.2cm}
	\caption{Illustration of the downlink MU-SemDaCom system.}
	\label{figure_system_diagram}
\end{figure}	
	
As shown in Fig. \ref{figure_system_diagram}, we consider a MU-MISO broadcast system, where a multi-antenna BS serves multiple single-antenna users with the same frequency band. The users can be categorized into two classes: DUs that decode received signals for source data reconstruction, and SUs that process these signals to accomplish semantic tasks.
\begin{figure*}[ht]
	\centering
	\includegraphics[width=0.8\linewidth]{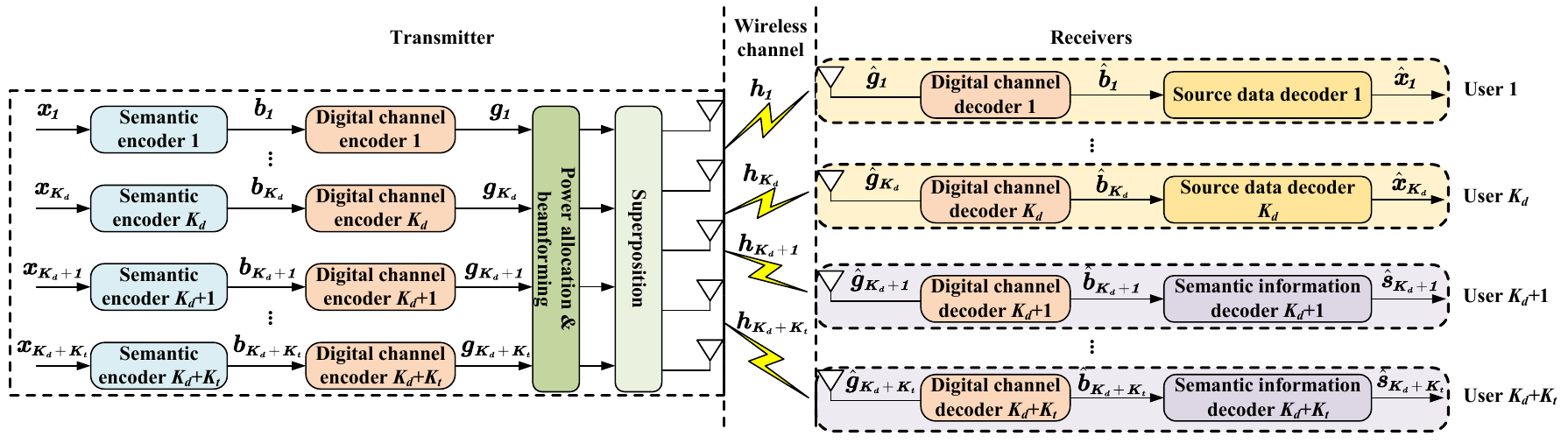}
	\vspace{-0.2cm}
	\caption{The MU-ASCC framework for the MU-SemDaCom system.}
	\label{figure_system_overview}
\end{figure*}
The system framework is shown in Fig. \ref{figure_system_overview} and the functionalities of each module are introduced as follows. 
\subsubsection{Transmitter}
At the transmitter, the BS is equipped with $N_t$ transmitting antennas. 
For each user $i\in\mathcal{K}$, it applies a semantic encoder to extract and compress semantic features, followed by a digital channel encoder that incorporates error protection against the channel noises and interferences. 
Subsequently, after power allocation and beamforming, the superposition coding scheme is applied to perform signal superposition across all users, generating a composite signal that is broadcast over wireless channels \cite{cover1972broadcast}. 
\subsubsection{Receiver}
At the receiver, each user first applies digital channel decoding to recover transmitted bit streams. Then, the DUs, indexed by $\mathcal{K}_d=\{1,2,...,K_d\}$, employ source data decoders to reconstruct source data, while the SUs, indexed by $\mathcal{K}_t=\{K_d+1,K_d+2,...,K_d+K_t\}$, utilize semantic decoders to execute semantic tasks. The total user number is $K=K_d+K_t$ and the set of all users is $\mathcal{K}=\mathcal{K}_d\cup\mathcal{K}_t$.
\vspace{-0.5cm} 
\subsection{Semantic Source Coding}
In this subsection, we introduce the semantic source coding schemes for DUs and SUs. 
\subsubsection{DUs} During the source encoding for user $i\in\mathcal{K}_d$, a semantic encoder is employed to compress the source data $\bm{x}_i\in\mathbb{R}^{d_X}$ into a bit stream $\bm{b}_i\in\{0,1\}^{B_i}$. $d_X$ is the dimension of $\bm{x}_i$ and $B_i$ is the length of $\bm{b}_i$. Specifically, as shown in Fig. \ref{figure_source_encoder}, the source data $\bm{x}_i$, carrying unknown semantic information $\bm{s}_i$, is processed through the DNN-based feature extraction function
\vspace{-0.3cm} 
\begin{align}
	\bm{y}_i=F_{\bm{\phi}_i}(\bm{x}_i),
\end{align}
 where $\bm{\phi}_i$ denotes the DNN parameters, $\bm{y}_i\in\mathbb{R}^{d_{Y_i}}$ is a continuous feature vector. Then, $\bm{y}_i$ is quantized as $\tilde{\bm{y}}_i\in\mathbb{R}^{d_Y}$ using the uniform scalar quantization \cite{balle2018variational}. Next, employing lossless entropy encoding methods (e.g., arithmetic encoding \cite{rissanen1979arithmetic}), $\tilde{\bm{y}}_i$ is compressed into the bit stream $\bm{b}_i$ with length $B_i$ and the expected source coding rate is $R_{s,i}=\mathbb{E}_{\bm{x}_i}\{B_i\}$. 

Upon receiving the recovered bit stream $\hat{\bm{b}}_i\in\{0,1\}^{B_i}$, the data source decoder $i$ reconstructs the original source data as $\hat{\bm{x}}_i\in\mathbb{R}^{d_{X_i}}$ through a two stage process as shown in Fig. \ref{figure_data_source_decoder}. First, the bit stream $\hat{\bm{b}}_i$ is decoded into a feature vector $\hat{\bm{y}}_i\in\mathbb{R}^{d_{Y_i}}$. Then, $\hat{\bm{y}}_i$ is processed by the DNN-based data recovery function $G_{\bm{\theta}_i}$ (parameterized by $\bm{\theta}_i$) to generate the final output $\hat{\bm{x}}_i\in\mathbb{R}^{d_{X_i}}$. 

\subsubsection{SUs} 
\begin{figure}[htbp]
\centering
\subfloat[]{
\includegraphics[width=0.8\columnwidth]{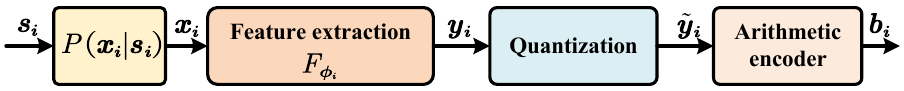}\label{figure_source_encoder}
}\vspace{-0.4cm} \\
\subfloat[]{
\includegraphics[width=.45\columnwidth]{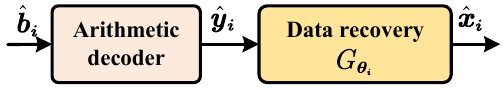}
\label{figure_data_source_decoder}
}
\subfloat[]{
\includegraphics[width=.45\columnwidth]{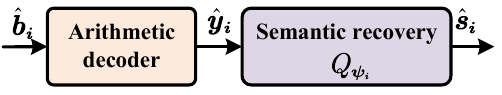}
\label{figure_task_source_decoder}
}
\vspace{-0.2cm}
\caption{Architectures of the semantic encoder, source data decoder and semantic information decoder. In Fig. \ref{figure_source_encoder}, $i\in\mathcal{K}_d\cup\mathcal{K}_t$. In Fig. \ref{figure_data_source_decoder}, $i\in\mathcal{K}_d$. In Fig. \ref{figure_task_source_decoder}, $i\in\mathcal{K}_t$.}
\label{figure_encoder_decoder}
\end{figure}

For the SU $i\in\mathcal{K}_t$, the source encoding, digital channel decoding and source decoding operations to obtain $\hat{\bm{y}}_i$ follow the same procedures as for DUs, as illustrated in Fig. \ref{figure_system_overview} and Fig. \ref{figure_encoder_decoder}. The decoded feature vector $\hat{\bm{y}}_i$ is then processed by the DNN-based semantic recovery function $Q_{\bm{\psi}_i}$ with network parameters $\bm{\psi}_i$ to reconstruct the semantic information as $\hat{\bm{s}}_i\in\mathbb{R}^{d_{S_i}}$. 
\subsubsection{Training Details}
In this paper, we consider a typical image transmission and classification scenario where the source data corresponds to images and semantic information is defined as their classification labels. The source coding DNNs for data reconstruction are trained over error-free channels, following the principle of rate-distortion theory \cite{balle2018variational}. This training process aims to determine the minimal source coding rate required to achieve the minimal data reconstruction distortion. The data distortion is measured by the MS-SSIM metric \cite{dai2022nonlinear}
\begin{align}
\mathcal{D}_o = \mathbb{E}_{\bm{x}}\left\{d_o(\bm{x},\hat{\bm{x}}) \right\},
\end{align}
where $d_o(\bm{x},\hat{\bm{x}}) $ computes the $1-$MS-SSIM value \cite{dai2022nonlinear}.
To overcome the non-differentiability of quantization operations, we employ a uniform noise with zero mean and unit radius to approximate the discrete quantization process \cite{balle2018variational}. The source coding DNNs for data reconstruction are optimized to balance the rate-distortion trade-off
\begin{align}
(\bm{\phi},\bm{\theta}) = \arg\min_{	(\bm{\phi},\bm{\theta})}R_s+\lambda \mathcal{D}_o,
\label{eqn_rd_train}
\end{align}
where $R_s$ is the source coding rate, and $\lambda$ is hyperparameter controlling the rate-distortion balance. By changing $\lambda$, we can obtain multiple source coding DNNs with different rate-distortion performances. 

The source coding DNNs for semantic task execution are also trained under perfect transmission conditions, leveraging the feature extractor $F_{\bm{\phi}}$ obtained from (\ref{eqn_rd_train}). Specifically, we first train an image classification DNN $E_{\bm{\vartheta}}$ where $\bm{\vartheta}$ denotes the DNN parameters and the loss function is the classification cross entropy.
Then, $Q_{\bm{\psi}}$ is constructed by cascading $G_{\bm{\theta}}$ with $E_{\bm{\vartheta}}$  \cite{huang2023joint}. The combined network is fine-tuned to minimize the cross entropy loss
\begin{align}
\bm{\psi}=\arg\min_{(\bm{\theta},\bm{\vartheta})}\mathbb{E}_{\bm{x}}\{L_{\text{CE}}(\bm{s},\hat{\bm{s}})\},	
\end{align}
where $L_{\text{CE}}(\bm{s},\hat{\bm{s}})$ measures the cross entropy between the true label $\bm{s}$ and the predicted label $\hat{\bm{s}}$.
\vspace{-0.3cm} 
\subsection{Finite Blocklength Transmissions}
In this subsection, we introduce the finite blocklength transmission process of the MU-SemDaCom system. To protect the data bits $\bm{b}_i$ against channel errors during transmission, the digital channel encoder $i$ encodes $\bm{b}_i$ as a complex symbol vector $\bm{g}_i\in\mathbb{C}^{d_{G_i}}$ with $\mathbb{E}_{\bm{x}_i}\{\frac{1}{d_{G_i}}\bm{g}_i^H\bm{g}_i\}=1$, where $d_{G_i}$ is the dimension of $\bm{g}_i$ representing the number of channel uses to transmit $\bm{x}_i$. Specifically, we consider a $(N_i,L)$ block channel code with channel coding rate $R_{c,i}=\frac{N_i}{L}$, which consists of a channel encoder $\mathcal{C}_i$ and a channel decoder $\mathcal{C}_i^{-1}$. $N_i$ is the length of the message bits and $L$ is the blocklength. First, $\bm{b}_i$ is divided into $\lceil \frac{B_i}{N_i} \rceil$ equal-length packets and each packet has length $N_i$. Next, these packets are encoded into complex-valued codewords with length $L$ by using the same $\mathcal{C}_i$ and concatenating the codewords $\bm{g}_i$. Accordingly, the average number of channel uses to transmit source data is $\mathbb{E}_{\bm{x}_i}\{d_{G_i}\}=\mathbb{E}_{\bm{x}_i}\{\lceil \frac{B_i}{N_i} \rceil L\}$. In practical wireless communication systems where $N_i$ is typically much smaller than the information bit length $B_i$, $\mathbb{E}_{\bm{x}_i}\{d_{G_i}\}$ can be approximated by $\frac{R_{s,i}}{R_{c,i}}$.
The BS employs superposition coding to simultaneously broadcast composite signals to all users by allocating power $p_i$ and applying unit beamforming vector $\bm{w}_i\in\mathbb{C}^{N_t}$ for each user $i\in\mathcal{K}$ \cite{cover1972broadcast}. Let $g_i^{(t)}$ be the $t$-th symbol in $\bm{g}_i$ being transmitted. At the $t$-th symbol period, the superposed signal is expressed by
\begin{equation}
	\bm{u}^{(t)}=\sum_{i\in\mathcal{K}}\sqrt{p_i}\bm{w}_ig_i^{(t)},
	\label{eqn_superpose}
\end{equation}
$t=1,2,...,d_{G_i}$.
 
The received signal of user $i$ at the $t$-th channel use is given by
\begin{equation}
	\hat{g}_i^{(t)}=\sqrt{p_i}\bm{h}_i^H\bm{w}_ig_i^{(t)}+\sum_{j\in\mathcal{K}\backslash\{i\}}\sqrt{p_j}\bm{h}_i^H\bm{w}_jg_j^{(t)}+n_i^{(t)},
	\label{eqn_super_i_recv}
\end{equation}
where $\bm{h}_i\in\mathbb{C}^{N_t}$ is the channel coefficient from the BS to user $i$ and $n_i^{(t)}$ is the independent and identically distributed (i.i.d.) circularly symmetric complex Gaussian (CSCG) noise with mean zero and variance $\sigma_i^2$. 
We consider the slow fading scenario, where $\bm{h}_i$ remains constant over image transmissions and is known at the transmitter and receiver side. The SINR is expressed as
\begin{equation}
	\gamma_i=\frac{p_i|\bm{h}_i^H\bm{w}_i|^2}{\sum_{j\in \mathcal{K}\backslash\{i\}}p_j|\bm{h}_i^H\bm{w}_j|^2+\sigma_i^2}.
	\label{eqn_sinr_def}
\end{equation}

At the receiver side, each user $i$ utilizes the channel decoder $\mathcal{C}^{-1}_i$ to decode the received signal $\hat{\bm{g}}_i$ into the bit stream $\hat{\bm{b}}_i\in\{0,1\}^{B_i}$. According to the finite blocklength transmission theory, the average packet error probability can be approximated by \cite{polyanskiy2010channel}
\begin{equation}
	\rho_i=Q\left(\frac{\sqrt{L}(\log_2(1+\gamma_i)-R_{c,i})}{\sqrt{\left(1-\frac{1}{(1+\gamma_i)^2}\right)\log_2^2e}}\right),
	\label{eqn_finite_rho}
\end{equation}
where $Q(x) = \frac{1}{\sqrt{2\pi}}\int_{x}^{\infty}e^{-\frac{t^2}{2}}dt$.
  \vspace{-0.3 cm}
\subsection{Distortion Evaluations}
This subsection introduces the E2E distortion evaluations for both the data reconstruction and semantic task execution. 
\subsubsection{E2E Distortion for DUs} In contrast to \cite{li2025adaptive} where mean squared error (MSE) is adopted, in this paper, we employ MS-SSIM as the data reconstruction metric since it is better aligned with perceptual quality assessments of reconstructed images \cite{dai2022nonlinear}. The average E2E distortion for DU $i$ is affected by the source distortion from source coding and transmission errors, which can be expressed by
\begin{align}
	\mathcal{D}_{o,i}=\mathbb{E}_{\bm{x}_i,\bm{n}_i}\left\{d_o(\bm{x}_i,\hat{\bm{x}}_i)\right\},
	\label{eqn_Doi_ssim} 
\end{align}
where $\bm{n}_i=[n_i^{(1)},n_i^{(2)},...,n_i^{(d_{G_i})}]$ denotes the noise vector.
\subsubsection{E2E Distortion for SUs}
For semantic distortion evaluation, we employ the Hamming distortion metric
\begin{equation}
	d_s(\bm{s}_i, \hat{\bm{s}}_i) = 
	\begin{cases} 
		0, & \text{if } \bm{s}_i = \hat{\bm{s}}_i, \\
		1, & \text{if } \bm{s}_i \neq \hat{\bm{s}}_i,
	\end{cases}
	\label{eqn_hamming}
\end{equation}
and the corresponding E2E semantic distortion is given by
\begin{equation}
	\mathcal{D}_{s,i}=\mathbb{E}_{\bm{x}_i,\bm{n}_i}\left\{d_s(\bm{s}_i,\hat{\bm{s}}_i)\right\}.
	\label{eqn_Dsi}
\end{equation}
\section{Problem Formulation}
\label{section_problem_formulation}
This section derives the analytical expressions for data and semantic E2E distortions and formulates the optimization problem for the proposed MU-ASCC scheme. 
\vspace{-0.4cm} 
\subsection{Distortion Modeling}
This subsection builds up analytical E2E distortion models for data reconstruction and semantic task execution.
The E2E distortions in (\ref{eqn_Doi_ssim}) and (\ref{eqn_Dsi}) exhibit complex dependencies on high-dimensional DNN parameters and channel conditions, making their analytical models difficult to obtain. To overcome this analytical intractability, following our preliminary work \cite{li2025adaptive}, we employ data regression methods to approximate the E2E distortions through logistic functions as detailed in the sequel.
\begin{figure*}[htbp]
\centering
\subfloat[]{
\includegraphics[width=.3\linewidth]{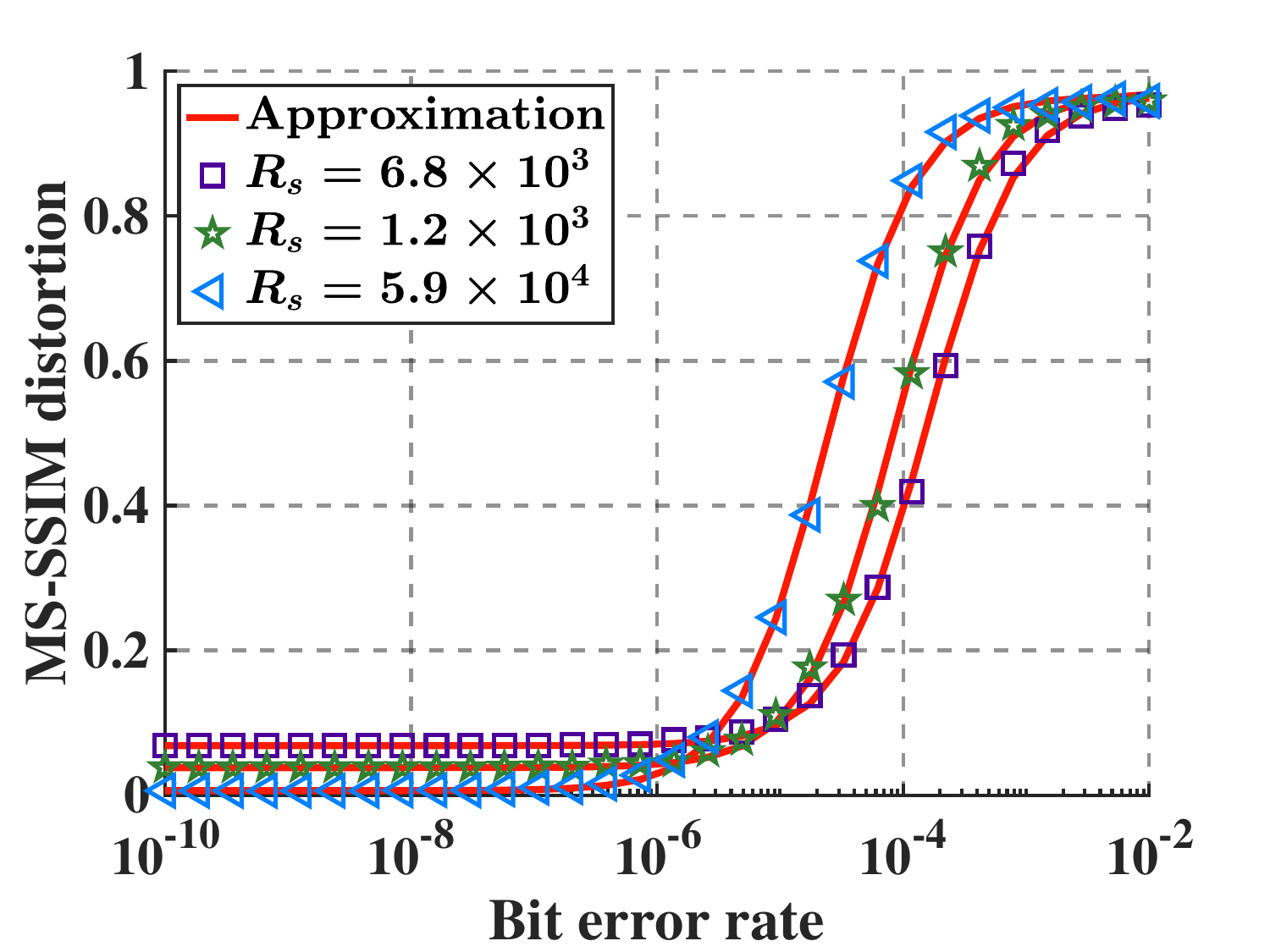}\label{figure_do_ssim_e2e_fix_rs}
}
\subfloat[]{
\includegraphics[width=.3\linewidth]{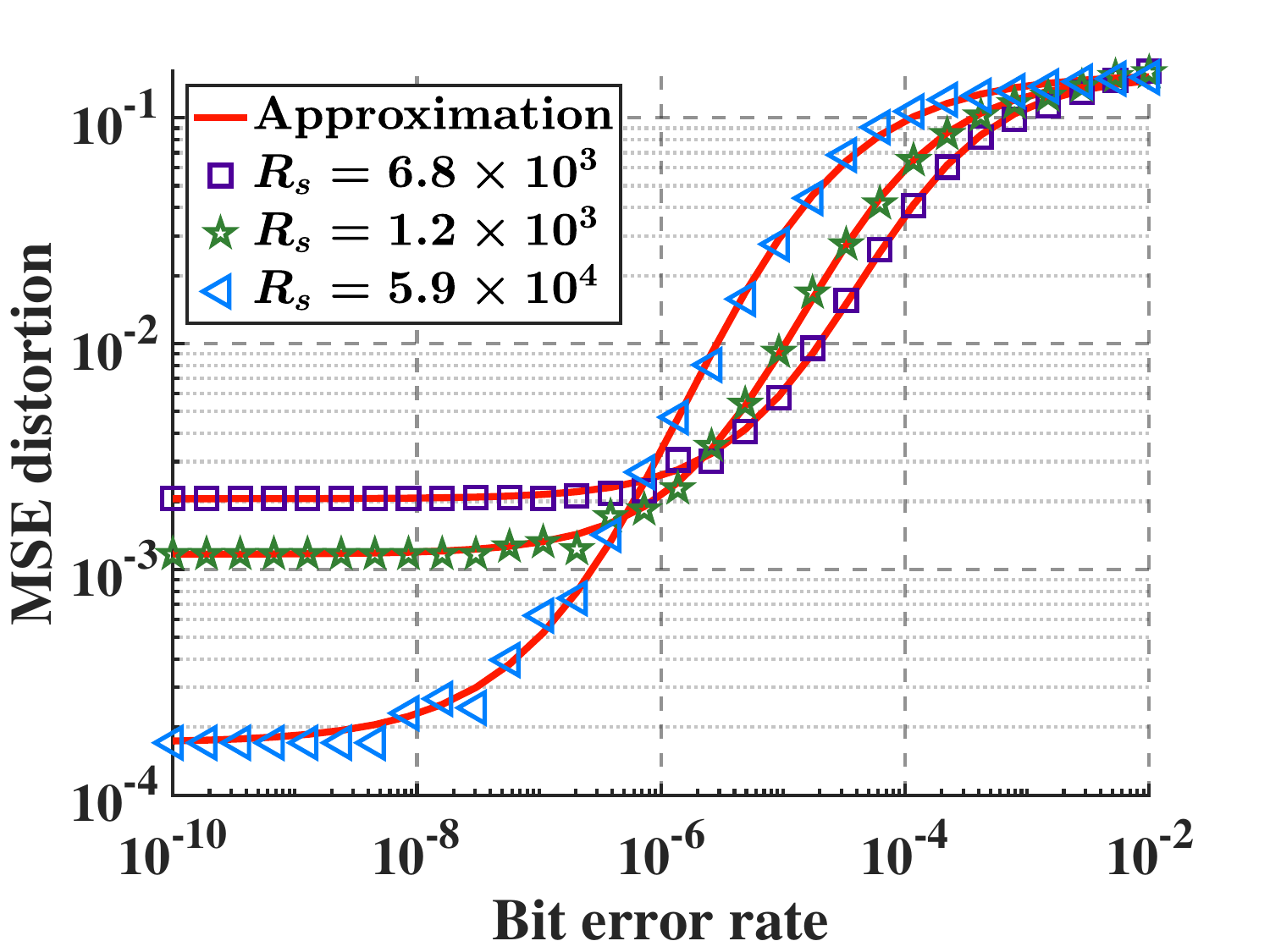}
\label{figure_do_logmse_e2e_fix_rs}
}
\subfloat[]{
\includegraphics[width=.3\linewidth]{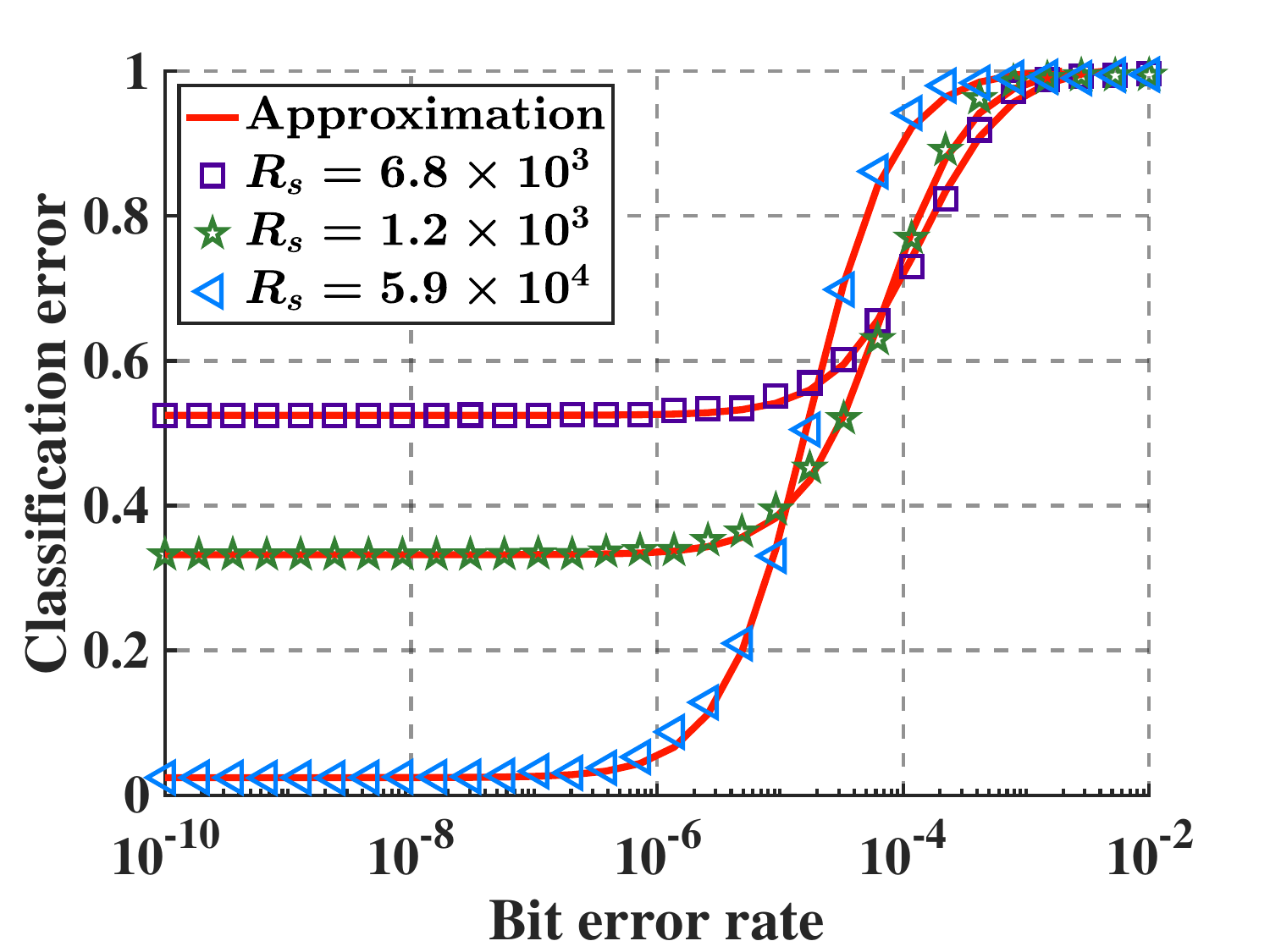}
\label{figure_ds_e2e_fix_rs}
}
\vspace{-0.2cm}
\caption{Comparison of the simulated results with logistic regression for approximating the average MS-SSIM distortion, MSE distortion and classification error over the CUB-200-2011 dataset. \(R_s\) is the source coding rate of the corresponding DNN model. The simulation settings, e.g., DNN architectures, are the same as the ones in the simulation section.}
\label{figure_do_e2e_fix_rs}
\end{figure*}
\begin{figure*}[htbp]
\centering
\subfloat[]{
\includegraphics[width=.3\linewidth]{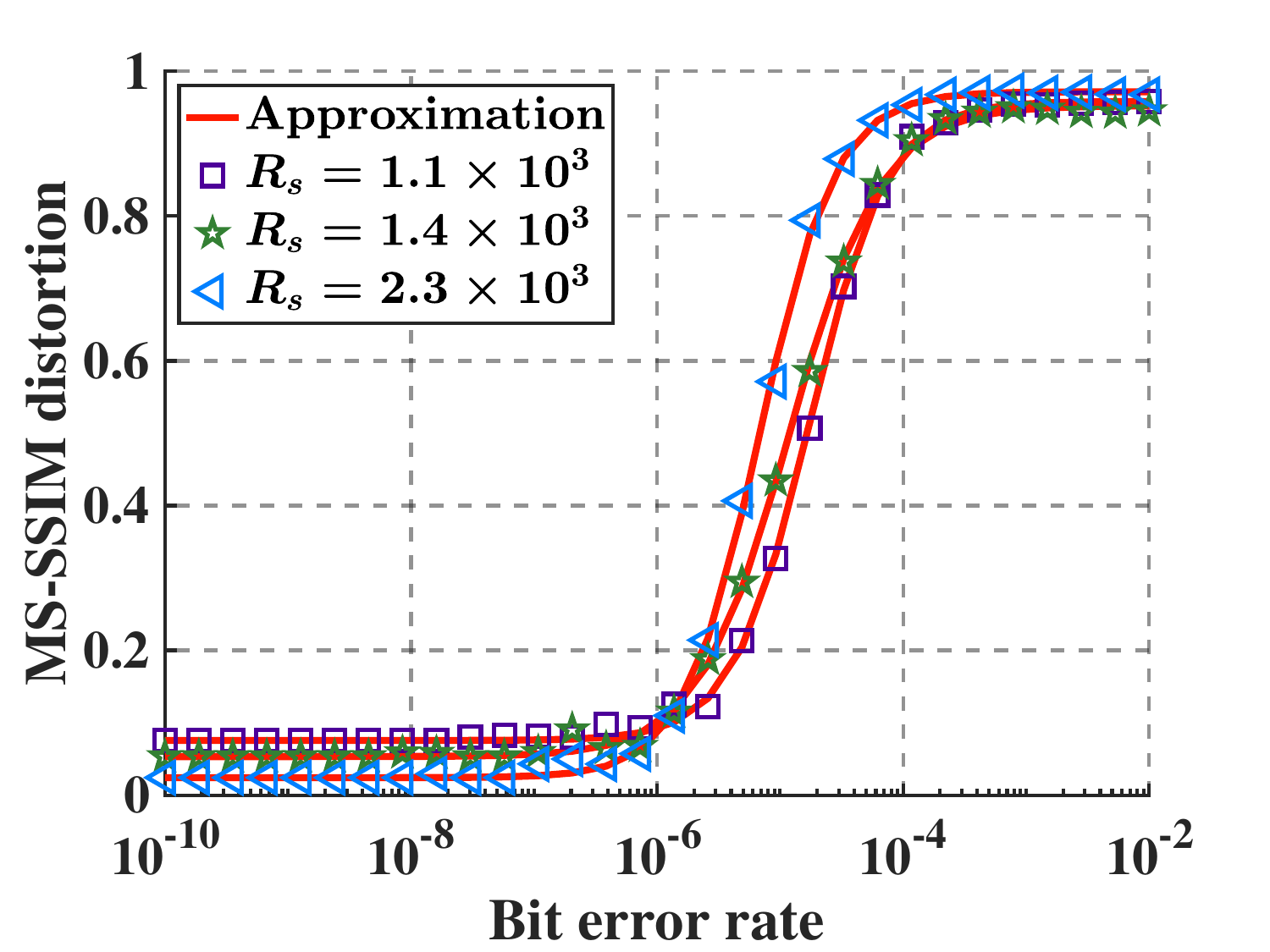}\label{figure_do_ssim_e2e_fix_rs_cifar}
}
\subfloat[]{
\includegraphics[width=.3\linewidth]{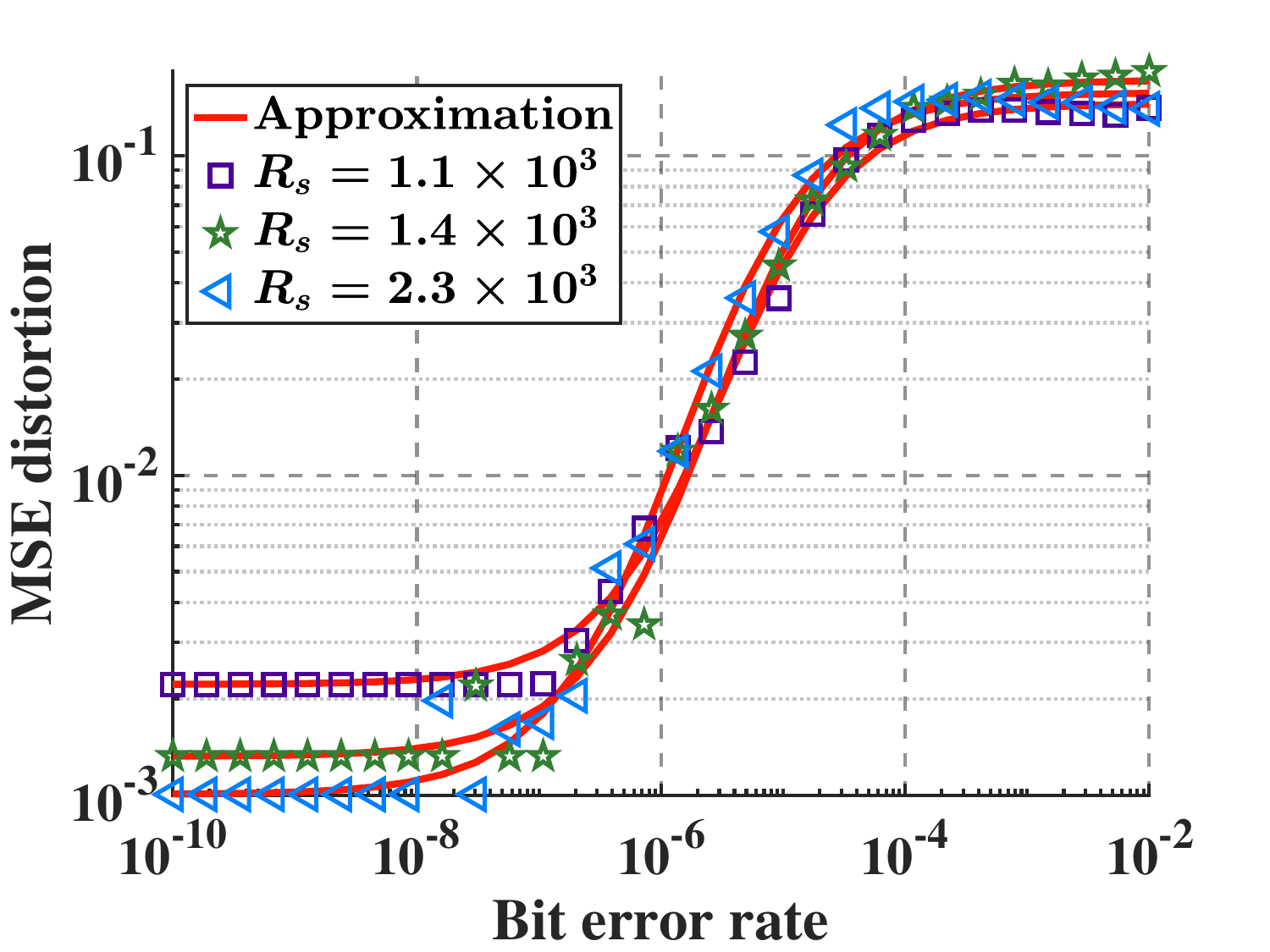}
\label{figure_do_logmse_e2e_fix_rs_cifar}
}
\subfloat[]{
\includegraphics[width=.3\linewidth]{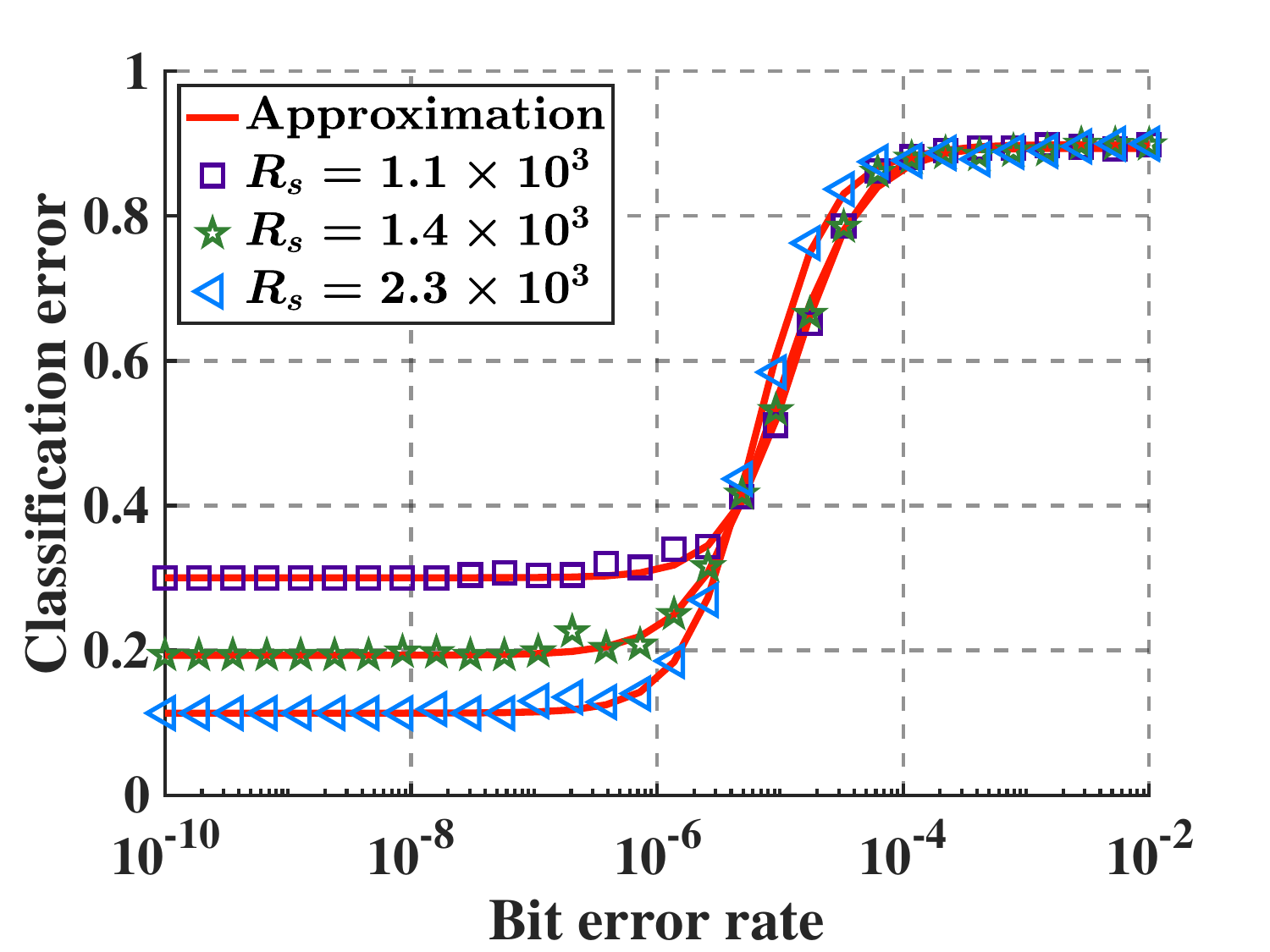}
\label{figure_ds_e2e_fix_rs_cifar}
}
\vspace{-0.2cm}
\caption{Comparison of the simulated results with logistic regression for approximating the average MS-SSIM distortion, MSE distortion and classification error over the CIFAR-10 dataset.}
\label{figure_do_e2e_fix_rs_cifar}
\end{figure*}

\subsubsection{E2E Distortion for DUs}
Specifically, for DUs, we first train $N_o$ DNN models for data recovery and construct a lookup table \(\mathcal{M}_d = \{R_{o,s}^{n},F_{\bm{\phi}_o^{n}}, G_{\bm{\theta}^{n}}\}_{n=1}^{N_o}\), where \(R_{o,s}^{n}\) denotes the source coding rate of \(F_{\bm{\phi}_o^{n}}\). 
The models in $\mathcal{M}_d$ are sorted in ascending order of their source coding rates, i.e, $R_{o,s}^n<R_{o,s}^{n+1}$ for $n=1,2,...,N_o-1$. 
For the $n$-th model in $\mathcal{M}_d$, let $d_{o,s}^n$ denote the average data distortion caused by source compression, which is determined by the source coding rate $R_{o,s}^n$ and is measured by $d_o$ under error-free transmissions. 
For each model in $\mathcal{M}_d$ with source coding rate $R_{o,s}^n$, we approximate the E2E distortion for DU $i$ measured by the $d_o$ metirc as the logistic function
\begin{equation}
        \label{eqn_do_logis}
	\tilde{\mathcal{D}}_{o,i}(R_{o,s}^{n},{\rho}_{b,i})\approx d_{o,s}^{n}+\frac{d_{o,c}^{n}}{1+e^{-a_{o,1}^{n}({\tilde{\rho}}_{b,i}-a_{o,0}^{n})}},
\end{equation}
Here, ${\tilde{\rho}}_{b,i}$ is the base-10 logarithm of the BER for user $i$, and $d_{o,c}^n$, $a_{o,1}^n$, $a_{o,0}^n$ are the logistic parameters for the $n$-th model in $\mathcal{M}_d$. The second term in (\ref{eqn_do_logis}) represents the distortion increment induced by channel errors, reaching its maximum value at the largest BER of DU $i$. 
The logistic parameters in (\ref{eqn_do_logis}) are estimated by the data regression method through minimizing the mean squared error between the predictions and observed data.

To validate the approximation (\ref{eqn_do_logis}), we calculate the E2E distortion by simulating channel errors through random bit flips in \(\bm{b}_i\) with probability \(\rho_{b,i}\), yielding the corrupted version $\hat{\bm{b}}_i$. We conduct the experiments on the widely studied Caltech-UCSD Birds 200 (CUB-200-2011)  \cite{wah2011caltech} and CIFAR-10 \cite{krizhevsky2009learning} datasets. 
Figs. \ref{figure_do_ssim_e2e_fix_rs}, \ref{figure_do_logmse_e2e_fix_rs}, \ref{figure_do_ssim_e2e_fix_rs_cifar}, and \ref{figure_do_logmse_e2e_fix_rs_cifar}, reveal that the logistic function (\ref{eqn_do_logis}) can accurately approximate the E2E distortion variations with respect to BER over different datasets.

\subsubsection{E2E Distortion for SUs}
The E2E distortion for SUs is modeled by applying the same logistic function approximation approach introduced for DUs, but replacing the data recovery DNNs with semantic task execution DNNs. Specifically, we train \(N_s\) DNN models for semantic task execution and build a lookup table \(\mathcal{M}_s = \{R_{s,s}^{n},F_{\bm{\phi}_s^{n}}, Q_{\bm{\psi}^{n}}\}_{n=1}^{N_s}\), where \(R_{s,s}^{n}\) is the source coding rate of \(F_{\bm{\phi}_s^{n}}\). 
The models in $\mathcal{M}_s$ are sorted in ascending order of their source coding rates, i.e, $R_{s,s}^n<R_{s,s}^{n+1}$ for $n=1,2,...,N_s-1$. 
For the $n$-th model in $\mathcal{M}_s$, $d_{s,s}^n$ denotes the average semantic distortion caused by source compresssion and semantic analysis, which is evaluated by the $d_s$ metric in (\ref{eqn_hamming}) and is determined by the source coding rate $R_{s,s}^n$ under error-free channel conditions. 
For the $n$-th model in $\mathcal{M}_s$, the E2E distortion for SU $i$ measured by the $d_s$ metirc is approximated as
\begin{align}
	\tilde{\mathcal{D}}_{s,i}(R_{s,s}^{n},{\rho}_{b,i})\approx d_{s,s}^{n}+\frac{d_{s,c}^{n}}{1+e^{-a_{s,1}^n(\tilde{\rho}_{b,i}-a_{s,0}^{n})}},
	\label{eqn_e2e_ds_rs_rho}
\end{align}
where the logistic parameters $d_{s,c}^{n}$, $a_{s,1}^{n}$ and $a_{s,0}^{n}$ are obtained via data regression through the minimum mean squared error criterion.
 The second term in (\ref{eqn_e2e_ds_rs_rho}) represents the additional semantic distortion caused by channel errors, attaining its maximum when SU $i$ has the largest BER. 
The experimental validations on the CUB-200-2011 and CIFAR-10 datasets in Figs. \ref{figure_ds_e2e_fix_rs} and \ref{figure_ds_e2e_fix_rs_cifar} demonstrate that (\ref{eqn_e2e_ds_rs_rho}) effectively characterizes the relationship between semantic distortion and BER.  
 \begin{remark}
According to the simulation results in Fig. \ref{figure_do_e2e_fix_rs}, we have the following observations: 
\begin{enumerate}
 \item MS-SSIM distortion is more robust than MSE distortion adopted in \cite{li2025adaptive}. For example, in Fig. \ref{figure_do_e2e_fix_rs}, at $R_s=5.9\times 10^4$, the performance degrades at $\text{BER}=10^{-6}$ under the MS-SSIM metric while degrades at $\text{BER}=10^{-9}$ under the MSE metric. This indicates that although some distortion occurs under the MSE metric, the image quality barely degrades under human perception, which makes the MS-SSIM metric more suitable to measure the data reconstruction performance.
 \item  Both the data and semantic distortion metrics exhibit varying BER tolerance with $R_s$. 
 For example, As illustrated in Fig. \ref{figure_ds_e2e_fix_rs}, models operating at lower $R_s$ values demonstrate enhanced robustness to channel errors, despite exhibiting higher source compression distortion. This is evidenced by the distortion threshold increasing from $10^{-6}$ to $10^{-5}$ when $R_s$ decreases from $5.9\times 10^4$ to $6.8\times 10^3$ in Fig. \ref{figure_ds_e2e_fix_rs}.
 This finding underscores the need for adaptive source coding based on channel conditions. 
\end{enumerate}
 \end{remark}
 
  \vspace{-0.3 cm}
 \subsection{Problem Formulation}
In this subsection, we formulate the optimization problem for the MU-ASCC framework to minimize the E2E distortions of the DUs and SUs. To analyze the effect of finite blocklength coding on the system, we approximate the decoding bit errors as i.i.d. Bernoulli random variables. Then, based on the average block error probability in (\ref{eqn_finite_rho}), the base-10 logarithm of BER can be approximately calculated as \cite{li2025adaptive}
\begin{equation}
	\label{eqn_log10rho}
	\tilde{\rho}_{b,i} \approx \log_{10}(\frac{1}{R_{c,i}L})+\log_{10}Q\left(\frac{\sqrt{L}\left(\log_2(1+\gamma_i)-R_{c,i}\right)}{\sqrt{\left(1-\frac{1}{(1+\gamma_i)^2}\right)\log_2^2(e)}}\right).
\end{equation}

By substituting (\ref{eqn_log10rho}) into (\ref{eqn_do_logis}) and (\ref{eqn_e2e_ds_rs_rho}), the E2E distortions in (\ref{eqn_Doi_ssim}) and (\ref{eqn_Dsi}) can be formulated as functions of the source coding rates $\{R_{s,i}\}$, channel coding rates $\{R_{c,i}\}$, beamforming $\{\bm{w}_i\}$, and transmitted power $\{p_i\}$. Our purpose is to jointly optimize these variables to minimize the weighted summation of the E2E distortions under the power budget and transmission delay constraints. In another word, the optimization problem can be formulated as
\begin{align}
	\text{(P1)}
	\min_{\substack{
			\{R_{s,i},R_{c,i},p_i,\bm{w}_i\}	}}
	\quad
	&\sum_{i\in\mathcal{K}_d}\beta_i\tilde{\mathcal{D}}_{o,i}+
	\sum_{j\in\mathcal{K}_t}\beta_j\tilde{\mathcal{D}}_{s,j}\\
	\text{s.t.} \quad
	\label{cst_data_rate}
	& R_{s,i}\in \mathcal{R}_o, \forall i\in\mathcal{K}_d,\\
	\label{cst_semantic_rate}
	& R_{s,i}\in \mathcal{R}_s, \forall i\in\mathcal{K}_t,\\
	&\label{cst_power}
	\sum_{i\in \mathcal{K}}p_i\le P_{max},\\
	&\label{cst_delay}
	\frac{R_{s,i}}{R_{c,i}}\le T_{i}, i\in\mathcal{K} \\
	&\label{cst_bf_norm}
	||\bm{w}_i||=1, \forall i\in\mathcal{K},,
\end{align}
where $\beta_i$ is a positive constant and denotes the weight of the E2E distortion for user $i$, $T_{i}$ denotes the maximum number of channel uses of user $i$, $\mathcal{R}_o=\{R_{o,s}^{1},...,R_{o,s}^{{N_o}}\}$ and $\mathcal{R}_s=\{R_{s,s}^{1},...,R_{s,s}^{N_s}\}$ represent the sets of source coding rates of the data reconstruction DNN models in $\mathcal{M}_d$ and semantic task execution DNN models in $\mathcal{M}_s$, respectively. Constraints (\ref{cst_data_rate}) and (\ref{cst_semantic_rate}) specify that the source encoders and decoders for DUs and SUs need to be selected from the pre-trained DNN models in $\mathcal{M}_d$ and $\mathcal{M}_s$, respectively. 
The two constraints are necessary for practical scenarios, where only a finite number of DNN models can be deployed\footnote{A key limitation for the MU-SemDaCom framework is the substantial memory overhead caused by storing the lookup tables $\mathcal{M}_d$ and $\mathcal{M}_s$. This can be addressed by employing the variable-rate source coding techniques \cite{kamisli2024variable}, which enable a single DNN model to support multiple rate-distortion operating points through a configurable scalar input parameter. Then the memory cost can be significantly reduced. }.
Constraint (\ref{cst_power}) guarantees that the transmission power remains within the power budget. Constraint (\ref{cst_delay}) limits the average transmission latency below the specified delay threshold for different users. Finally, constraint (\ref{cst_bf_norm}) imposes the unit norm requirement on the beamforming vectors.
Problem (P1) does not include an explicit channel capacity constraint because the objective function inherently penalizes cases where channel coding rates exceed capacities. When this occurs, as shown in Fig. \ref{figure_do_e2e_fix_rs}, the resulting channel errors drive the E2E distortion to its maximum value, making such solutions undesirable.
Problem (P1) is a \emph{mixed-integer nonlinear programming} (MINLP) problem \cite{boyd2004convex} with a complicated non-convex objective function. In addition, there are two major difficulties for the joint optimization: the coexistence of discrete and continuous optimizing variables and their strong coupling in the objective function.
\begin{remark}
In Problem (P1), the BER relationship is used to describe the performance of \emph{random coding}, which is an ideal coding scheme in the finite block length transmission \cite{polyanskiy2010channel}. Solving Problem (P1) leads to a performance bound of the MU-SemDaCom system. However, the proposed system and optimization formulation can be easily extended into scenarios with practical channel coding and modulations by utilizing their corresponding BER relationships \cite{li2025adaptive,hassani2014finite}. 
\end{remark}

  \vspace{-0.3 cm}
\section{Joint Rate, Power and Beamforming Optimization}
\label{section_joint_rate_power_beam}

In this section, we introduce the solution to the joint source-channel coding rate, power, and beamforming optimization in Problem (P1). 
  \vspace{-0.3 cm}
\subsection{Overview of the Algorithm}
\begin{figure}[h]
\centering
\includegraphics[width=0.8\linewidth]{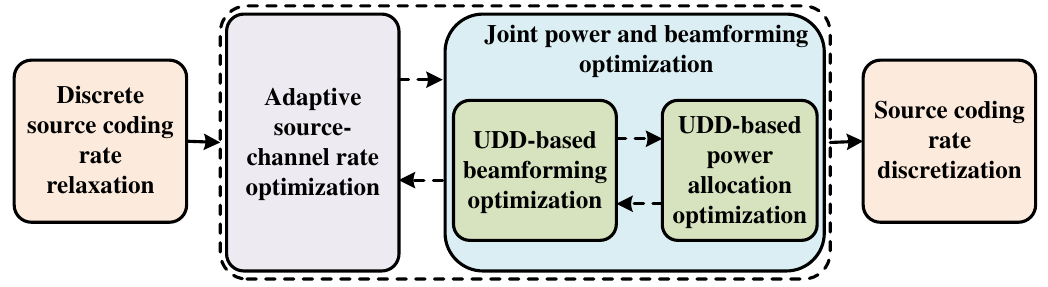}	
\vspace{-0.2cm}
\caption{Framework of the joint rate, power and beamforming optimization algorithm.}
\label{fig_opt}
\end{figure}

This subsection gives an overview of the JRPB algorithm. As illustrated in Fig. \ref{fig_opt}, we first relax the discrete source coding rates $\{R_{s,i}\}$ into continuous variables and transform Problem (P1) into a continuous optimization problem. Since power allocation and beamforming vectors determine the channel capacity of each user while source-channel coding rates control both the transmission latency and E2E distortion, we employ an AO framework to decompose the relaxed problem into two corresponding subproblems:
\begin{itemize}
\item \textbf{Adaptive source-channel rate optimization}: In this subproblem, we fix the power allocation and beamforming vectors $\{p_i,\bm{w}_i\}$, and optimize the source-channel coding rates $\{R_{s,i},R_{c,i}\}$ to minimize the weighted-sum E2E distoriton under the transmission delay constraints.
\item \textbf{Joint power and beamforming optimization}: In this subproblem, we fix the source-channel coding rates $\{R_{s,i},R_{c,i}\}$, and optimize the power allocation and beamforming vectors $\{p_i,\bm{w}_i\}$ to minimize the weighted-sum E2E distoriton under the power budget and unit beamforming constraints.
\end{itemize}
We first solve the two subproblems alternately until convergence, then discretize the continuous source coding rates in the converged solution to obtain the final solution to Problem (P1). 
The overall algorithm is shown in Fig. \ref{fig_opt}.

  \vspace{-0.3 cm}
\subsection{Adaptive Source-Channel Rate Optimization}
When the power and beamforming $\{p_i,\bm{w}_i\}$ are fixed, the objective value depends solely on $\{R_{s,i},R_{c,i}\}_{\mathcal{K}}$. In this case, the source and channel coding rates of each user do not affect the distortions of other users. This decoupling property allows us to decompose the weighted-sum distortion minimization in Problem (P1) into independent per-user distortion minimization problems, which can be expressed as follows
\begin{align}
	\text{({P2})}
	\min_{R_{s,i}, R_{c,i}} & \ \tilde{\mathcal{D}}_{k_i,i}(R_{s,i}, \tilde{\rho}_{b,i}) \\
	\text{s.t.} & \
	\label{cst_rate_delay}
	\frac{R_{s,i}}{R_{c,i}}\le T_{i},\\
	\label{cst_rate_set}
	& R_{k_i,s}^1\le R_{s,i}\le R_{k_i,s}^{N_{k_i}},
\end{align}
where $k_i$ is an indicator being $o$ if $i\in\mathcal{K}_d$ or $s$ if $i\in\mathcal{K}_t$, and the discrete source coding rate constraint (\ref{cst_data_rate}) or (\ref{cst_semantic_rate}) for user $i$ is relaxed as (\ref{cst_rate_set}). The parameter \(\tilde{\rho}_{b,i}\) is derived from (\ref{eqn_log10rho}).

As proved in \cite{huang2025d},  the solution to Problem (P2) is achieved when $\frac{R_{s,i}}{R_{c,i}}=T_i$. Thus Problem (P2) can be simplified as  the following single-variable optimization problem
\begin{align}
    \text{({P2.1})}
    \quad
    \min_{R_{s,i}} & \ \tilde{\mathcal{D}}_{k_i,i}(R_{s,i}, \tilde{\rho}_{b,i}) \\
    \text{s.t.} & \
    \label{cst_Rsi}
    \ R_{k_i,s}^1\le R_{s,i} \le R_{k_i,s}^{N_{k_i}},
\end{align}
where $\tilde{\rho}_{b,i}$ is obtained by replacing $R_{c,i}$ in (\ref{eqn_log10rho}) with $\frac{R_{s,i}}{T_{i}}$.
 
To solve Problem {(P2.1)}, one difficulty is that $\tilde{\mathcal{D}}_{k_i,i}$ is only defined on $\mathcal{R}_{k_i}\times \mathbb{R}$.
To extend the function domain of $\tilde{\mathcal{D}}_{k_i,i}$ to $\mathbb{R}\times\mathbb{R}$, we use linear interpolation technique to approximate the E2E distortion. When the source coding rate is not in $\mathcal{R}_{k_i}$, for $R_{s,i}$ satisfying $R_{k_i,s}^n\le R_{s,i}<R_{k_i,s}^{n+1}$, based on (\ref{eqn_do_logis}) and (\ref{eqn_e2e_ds_rs_rho}), $\tilde{\mathcal{D}}_{k_i,i}$ is approximated as
\vspace{-0.3cm} 
\begin{align}
\label{eqn_conti_D}
\tilde{\mathcal{D}}_{k_i,i}(R_{s,i},\tilde{\rho}_{b,i})\approx \tilde{d}_{k_i,s}+\frac{\tilde{d}_{k_i,c}}{1+e^{-\tilde{a}_{k_i,1}(\tilde{\rho}_{b,i}-\tilde{a}_{k_i,0})}},
\end{align}
where 
\begin{align}
\label{eqn_tilde_d_os}
\tilde{d}_{k_i,s} = d_{k_i,s}^n+\lambda_i (d_{k_i,s}^{n+1}-d_{s}^n),\\
\tilde{d}_{k_i,c} = d_{k_i,c}^n+\lambda_i (d_{k_i,c}^{n+1}-d_{k_i,c}^n),\\
\tilde{a}_{k_i,1} = a_{k_i,1}^n+\lambda_i (a_{k_i,1}^{n+1}-a_{k_i,1}^n),\\
\label{eqn_tilde_a_o0}
\tilde{a}_{k_i,0} = a_{k_i,0}^n+\lambda_i (a_{k_i,0}^{n+1}-a_{k_i,0}^n),
\end{align}
with $\lambda_i = \frac{R_{s,i}-R_{k_i,s}^n}{R_{k_i,s}^{n+1}-R_{k_i,s}^n}$.

Using the approximation in (\ref{eqn_conti_D}), Problem {(P2.1)} reduces to minimizing a continuous function over a bounded interval. We solve it via subgradient descent. At iteration $n$, the update is
$
R_{s,i}^{(n+1)} = R_{s,i}^{(n)}-\alpha_nd^{(n)},	
$
where $\alpha_n$ is the stepsize and $d^{(n)}$ is a subgradient of $\tilde{\mathcal{D}}_{k_i,i}(R_{s,i},\tilde{\rho}_i)$ at $R_{s,i}^{(n)}$. Since $\tilde{\mathcal{D}}_{k_i,i}(R_{s,i},\tilde{\rho}_i)$ is differentiable except at discrete points in $\mathcal{R}_{k_i}$, we set $d^{(n)}$ as the gradient when $R_{s,i}\in [R_{k_i,s}^1,\bar{R}_{s,i}]\cap \mathcal{R}_{k_i}^c$, where $\mathcal{R}_{k_i}^c$ is the complement of $\mathcal{R}_{k_i}$ on $\mathbb{R}$. When $R_{s,i}\in\mathcal{R}_{k_i}$, 
the subdifferential $\partial\tilde{\mathcal{D}}_{k_i,i}(R_{s,i},\tilde{\rho}_i)=[d^{-}(R_{s,i}),d^{+}(R_{s,i})]$, where $d^{-}(R_{s,i})$ is the left derivative and $d^{+}(R_{s,i})$ is the right derivative at $R_{s,i}$. We set 
\begin{align}
\label{eqn_subgrad}
d^{(n)} = 
\begin{cases}
d^{+}(R_{s,i}), & \text{if } R_{s,i}=R_{k_i,s}^1,\\
d^{-}(R_{s,i}), &\text{if } R_{s,i}=R_{k_i,s}^N, \\
 \frac{d^{-}(R_{s,i})+d^{+}(R_{s,i})}{2}, &\text{otherwise}.
\end{cases}	
\end{align}
The stepsize is determined via backtracking line search \cite{boyd2004convex} and the algorithm is summarized in Algorithm \ref{alg_rate}.

\begin{algorithm}[hbt]
\caption{Adaptive Source-Channel  Rate Optimization Algorithm.}
\label{alg_rate}
\begin{algorithmic}[1]
\Require $\{\bm{h}_i,\sigma_i,p_i,\bm{w}_i\}$,
	$T_{i}$.
\Ensure {$\{R_{s,i}^*,R_{c,i}^*\}$.}
\State \textbf{For} {each user $i$ in the $K$ users}
\State \quad Set the iteration number $n=1$ and the starting point $R_{s,i}^{(n)}=R_{k_i,s}^1$.

\State \quad \textbf{Repeat}
\State \quad \quad Compute the subgradient $d^{(n)}$ as the derivative of $\tilde{\mathcal{D}}_i$ if $R_{s,i}\notin\mathcal{R}_{k_i}$. Otherwise, compute $d^{(n)}$ based on (\ref{eqn_subgrad}).
\State \quad \quad Compute the stepsize $\alpha_n$ using the backtracking line search \cite{boyd2004convex}.
\State \quad \quad Update $R_{s,i}^{(n+1)} = R_{s,i}^{(n)}+\alpha_nd^{(n)}$.
\State \quad \quad Update $n=n+1$.
\State \quad \textbf{Until} The fractional decrease of the objective value is below a threshold $\epsilon$.
\State \quad Set $R_{s,i}^*=R_{s,i}^{(n)}$ and compute $R_{c,i}^*=\frac{R_{s,i}^*}{T_{i}}$.
\State \textbf{End for}
\end{algorithmic}
 \end{algorithm}
 \vspace{-0.5cm} 
\subsection{Joint Power and Beamforming Optimization}
When the source and channel coding rates are fixed, the joint power and beamforming optimization subproblem is
 \begin{align}
		\text{(P3)}
		\min_{\substack{
				\{p_i,\bm{w}_i\}
			}}
		\quad
		&\sum_{i\in\mathcal{K}}\beta_i\tilde{\mathcal{D}}_{k_i,i}(R_{s,i},\tilde{\rho}_{b,i})\\
		\text{s.t.} \quad
		&\nonumber
		(\ref{cst_power}),(\ref{cst_bf_norm}).
\end{align}
Here, power and beamforming influence the transmission distortion through their impact on BER. When the blocklength and channel coding rate are fixed, BER appears to be solely determined by SINR, as depicted by (\ref{eqn_log10rho}). 
The main challenge is the interdependence among users: the power and beamforming of each user influence the SINR and distortions of other users.

 To decouple these interdependencies, we use UDD theory to transform the downlink problem into a dual uplink problem. 
 The virtual uplink system has $K$ single-antenna transmitting users with the same grouping and indexing as the downlink system. A receiver with $N_t$ antennas performs signal reception and data/semantic decoding for each user. The power, unit beamforming vector, and channel coefficient of user $i$ are denoted as $q_i$, $\bm{w}_i^u$, and $\bar{\bm{h}}_i=\frac{\bm{h}_i}{\sigma_i^2}$, respectively. The AWGN noise is $\bm{n}\sim\mathcal{CN}(0,\bm{I}_{N_t})$. The total power is constrained by $P_{max}$. The uplink SINR for user $i$ is computed as
 \begin{align}
    \label{eqn_up_sinr}
    \gamma_i^u = \frac{q_i|\bar{\bm{h}}_i^H\bm{w}_i^u|^2}{\sum_{j\in\mathcal{K}/\{i\}}q_{j}|\bar{\bm{h}}_{j}^H\bm{w}_{i}^u|^2+1}.
\end{align}

Let the power allocation of the downlink system be represented as \(\bm{p} = [p_1, p_2, \ldots, p_K]^{T}\) and the power allocation of the uplink system be represented as \(\bm{q} = [q_1, q_2, \ldots, q_K]^T\). According to \cite{rashid1998transmit}, when the power budgets of the two systems are the same, i.e., $\sum_{i=1}^K p_i=\sum_{i=1}^Kq_i=P_{max}$, we have \(\gamma_i = \gamma_i^u, \forall i\in\mathcal{K},\) when the uplink and downlink power allocations satisfy the following relationship
\begin{align}
	\label{eqn_q2p}
	&\bm{p}=\bm{\Psi}^{-1}\bm{1}_K,\\
	\label{eqn_p2q}
	&\bm{q} = \bm{\Phi}^{-1}\bm{1}_K,
\end{align}
with
\begin{align}
	[\bm{\Psi}]_{k,l}=\left\{\begin{aligned}
		&\frac{|\bm{\bar{h}}_k^H\bm{w}_k^u|^2}{\gamma_k^u},k=l,\\
		&|\bm{\bar{h}}_k^H\bm{w}_l^u|^2,k\ne l.
	\end{aligned}\right.
\end{align}
and
\begin{align}
	[\bm{\Phi}]_{k,l}=\left\{\begin{aligned}
		&\frac{|\bm{\bar{h}}_k^H\bm{w}_k|^2}{\gamma_k},k=l,\\
		&|\bm{\bar{h}}_l^H\bm{w}_k|^2,k\ne l.
    \end{aligned}\right.
\end{align}

Based on the above description for the virtual uplink system, the uplink joint power and beamforming problem is formulated as
    \begin{align}
    \label{eqn_up_obj}
		\text{(P4)}
		\min_{\substack{
				\{q_i,\bm{w}_i\}}}
		\quad
		&\sum_{i\in\mathcal{K}}\beta_i\tilde{\mathcal{D}}_{k_i,i}(R_{s,i},\tilde{\rho}_{b,i}^u)\\
		\text{s.t.} \quad
		&\label{cst_up_power}
        \sum_{i\in \mathcal{K}}q_i\le P_{max},\\
        &\label{cst_up_bf_norm}
        ||\bm{w}_i^u||=1, \forall i\in\mathcal{K}
		,
    \end{align}
    where $\tilde{\rho}_{b,i}^u$, $i\in\mathcal{K}$, is computed from (\ref{eqn_log10rho}) by replacing $\gamma_i$ as $\gamma_i^u$, (\ref{cst_up_power}) is the power constraint for the virtual uplink system, and (\ref{cst_up_bf_norm}) represents the uplink beamforming vector unit norm constraint. 
When Problem (P4) is solved, the solution to Problem (P3) can be obtained by the following proposition. 
\begin{proposition}
    \label{prop_p3p4}
    Denote the optimal solution to Problem {(P4)} as $\{\bm{w}_i^{u*},q_i^*\}$. Then the optimal solution to Problem {(P3)} $\{\bm{w}_i^*,p_i^*\}$ can be obtained by setting $\bm{w}_i^*=\bm{w}_i^{u*}$ and computing $\bm{p}^*=[p_1^*, p_2^*, \ldots, p_K^*]^{T}$ based on (\ref{eqn_q2p}).
\end{proposition}
\begin{IEEEproof}
Please see Appendix \ref{proof_prop_p3p4}.
\end{IEEEproof}

To solve Problem (P4), as shown in Fig. \ref{fig_opt}, we apply the AO method to decompose it into a beamforming optimization problem and a power allocation optimization problem. 
 \subsubsection{UDD-based Beamforming Optimization}

	According to (\ref{eqn_up_sinr}) and (\ref{eqn_up_obj}), when  \(\{q_i\}\) is fixed, the distortion for user $i$ only depends on $\bm{w}_i^u$. Minimizing the weighted-sum distortion is equivalent to individually minimizing the distortion for each user. Therefore, the beamforming optimization subproblem of Problem {(P4)} can be simplified as solving the following subproblem for each user $i$
     \begin{align}
        \text{({P5})}
        \quad
        \min_{\bm{w}_i\in\mathbb{C}^{N_t}} & \ \tilde{\mathcal{D}}_{k_i,i}(R_{s,i}, \tilde{\rho}_{b,i}^u) \\
        \text{s.t.} & \
        \ ||\bm{w}_i^u||=1.
    \end{align}
    It is easy to see that $\tilde{\mathcal{D}}_{k_i,i}$ monotonically decreases as $\gamma_i^u$ increases. Thus, the optimal solution for Problem {(P5)} is the beamforming that maximizes $\gamma_i^u$. When the uplink transmitting power $\{q_i\}$ is fixed, $\gamma_i^u$ is maximized by the MMSE beamforming \cite{he2021beamforming}. Therefore,  the optimal solution $\bm{w}_i^{u*}$ for Problem {(P5)} is the MMSE beamforming, expressed as
    \begin{align}
		\label{eqn_mmse}
		\bm{w}_i^{u*}=\frac{(\bm{I}_{N_t}+\sum_{i=1}^Uq_i\bar{\bm{h}}_i\bar{\bm{h}}_i^H)^{-1}\bar{\bm{h}}_i}{||(\bm{I}_{N_t}+\sum_{i=1}^Uq_i\bar{\bm{h}}_i\bar{\bm{h}}_i^H)^{-1}\bar{\bm{h}}_i||}.
    \end{align}
  \subsubsection{UDD-based Power Allocation Optimization}
    Given the beamforming vectors $\{\bm{w}_i^u\}$, the power allocation subproblem of {(P4)} is formulated as
    \begin{align}
		\text{(P6)}
		\min_{\substack{
				\{q_i\}
			}}
		\quad
        \label{obj_p8}
		&\sum_{i\in\mathcal{K}}\beta_i\tilde{\mathcal{D}}_{k_i,i}(R_{s,i},\tilde{\rho}_{b,i}^u)\\
		\text{s.t.} \quad \nonumber
		&(\ref{cst_up_power}).
    \end{align}
    Despite constraint (\ref{cst_up_power}) being linear with respect to $\{q_i\}$, Problem {(P6)} is non-convex since the coupling among the distortions of users and the complicated expression of the E2E distortions. We use the successive convex approximation (SCA) method to obtain a suboptimal solution for Problem {(P6)}.  Introducing slack variables $t_i$, $\hat{\rho}_i$, $g_i$, $\zeta_i$, and $\xi_i$, $i\in\mathcal{K}$, Problem {(P6)} is transformed into
    \begin{align}
		\text{(P6.1)}\nonumber\\
		\min_{\left\{\substack{
				q_i,t_i,\hat{\rho}_i,\\
                g_i,\zeta_i,\xi_i
			}\right\}}&
        \label{obj_sca0}
         \sum_{i\in\mathcal{K}}\beta_i\tilde{d}_{k_i,s}(R_{s,i})+\frac{\beta_i\tilde{d}_{k_i,c}(R_{s,i})}
        {1+{c}_{k_i,i}t_{i}}\\
		\text{s.t.} \quad &
         \label{cst_t}
        \log(t_i) +\tilde{a}_{k_i,1}(R_{s,i})\log_{10}Q\left(\frac{\hat{\rho}_i}{R_{c,i}L}\right) \le 0,\\
        \label{cst_hatrho}
        & \hat{\rho}_ig_i -\frac{\sqrt{L}}{\log_2e}(\log_2(1+\zeta_i)-R_{c,i})\le 0, \\
        \label{cst_gi}
        & 1-\frac{1}{(1+\xi_i)^2}-g_i^2 \le 0,\\
        \label{cst_zeta}
        & \sum_{j\ne i}
        \zeta_iq_{j}|\bar{\bm{h}}_{j}^H\bm{w}_{i}|^2+
        \zeta_i-
        q_i|\bar{\bm{h}}_i^H\bm{w}_i|^2\le0,\\
        \label{cst_xi}
        & -\sum_{j\ne i}
        \xi_iq_{j}|\bar{\bm{h}}_{j}^H\bm{w}_{i}|^2-
        \xi_i+
        q_i|\bar{\bm{h}}_i^H\bm{w}_i|^2\le0,\\
        \label{cst_bounds}
        & t_i\ge 0, q_i\ge 0, 0\le g_i\le 1, 0\le \zeta_i,\xi_i \le \bar{\gamma}_i^u, i\in\mathcal{K},\\
        &\nonumber (\ref{cst_up_power}),
    \end{align}
    where $c_{k_i,i}=e^{\tilde{a}_{k_i,1}(R_{s,i})\tilde{a}_{k_i,0}(R_{s,i})}$ is a positive constant when the rate allocation for all users are fixed and $\bar{\gamma}_{i}^u = P_{max}|\bm{\bar{h}}_i^H\bm{w}_i^u|^2$ is the SINR for user $i$ when $q_i=P_{max}$. 
    
    In Problem {(P6.1)}, even though the objective function is convex, all constraints, except for (\ref{cst_up_power}) and (\ref{cst_bounds}), are non-convex. 
    To deal with these non-convex constraints, the SCA method is employed to obtain a convex upper bound for the left-hand sides (LHSs) of the non-convex constraints.
    The following lemma uses the SCA method to obtain a convex estimate for the LHS of (\ref{cst_t}).
    \begin{lemma}
    \label{lemma_sca_logQ}
        Denote $\hat{Q}(x) = \log Q(x)$. Let $t_i^{(n)}$ and $\hat{\rho}_i^{(n)}$ be the feasible solution obtained from the $n$-th SCA iteration. The LHS of (\ref{cst_t}) has a convex upper bound, i.e.,
        \begin{align}
        \label{cst_t_sca}
            &\log(t_i) +\tilde{a}_{k_i,1}(R_{s,i})\left(\log_{10}(\frac{1}{R_{c,i}L})+\log_{10}Q\left(\hat{\rho}_i\right)\right) \nonumber\\
            & \le U(t_i^{(n)},\hat{\rho}_i^{(n)},t_i,\hat{\rho}_i), i\in\mathcal{K},
        \end{align}
        where 
        \begin{align}
            &U(t_i^{(n)},\hat{\rho}_i^{(n)},t_i,\hat{\rho}_i)=\frac{t_i}{t_i^{(n)}}+\log(t_i^{(n)})-1\nonumber\\
            & +\tilde{a}_{k_i,1}(R_{s,i})\left(\log_{10}(\frac{1}{R_{c,i}})+\tilde{Q}(\hat{\rho}_i^{(n)},\hat{\rho}_i)\right).
        \end{align}
        $\tilde{Q}$ is a linear function with respect to $\hat{\rho}_i$ expressed as
        \begin{align}
            \tilde{Q}(\hat{\rho}_i^{(n)},\hat{\rho}_i)=\left(\hat{Q}'(\hat{\rho}_i^{(n)})(\hat{\rho}_i-\hat{\rho}_i^{(n)})+\hat{Q}(\hat{\rho}_i)\right)\log_{10}e.
        \end{align}
        $\hat{Q}'$ is the derivative of $\hat{Q}$.
    \end{lemma}
    \begin{IEEEproof}
      Please see Appendix \ref{proof_lemma_sca_logQ}.
    \end{IEEEproof}
  \begin{algorithm}[hbt]
\caption{SCA-Based Uplink Power Allocation Algorithm for Problem {(P4)}}
\label{alg_power}
\begin{algorithmic}[1]
	\Require $\{\bar{\bm{h}}_i,R_{s,i},R_{c,i},\bm{w}_i^u,\beta_i\},P_{max}$.
	\Ensure $\{q_i^*\}$.
	\State Set iteration number $n=1$.
	\State Initialize the local points $\bm{\Theta}^{(n)}=\{t_i^{(n)}, \hat{\rho}_i^{(n)}, d_i^{(n)},g_i^{(n)},\zeta_i^{(n)},\xi_i^{(n)}, q_i^{(n)}\}$, for Problem {(P6.2)}.
	\State \textbf{Repeat}
	\State \quad Solve Problem {(P6.2)} at current local points $\bm{\Theta}^{(n)}$ using convex optimization toolbox, and obtain solution $\bm{\Theta}^{(n)*}$.
	\State \quad Update the local points as $\bm{\Theta}^{(n+1)}=\bm{\Theta}^{(n)*}$.
	\State \quad Update the iteration number $n=n+1$.
	\State \textbf{Until} \textnormal{the fractional decrease of the objective value of Problem {(P4)} is below a threshold $\epsilon_1$}.
	\State Obtain $\{q_i^*\}$ as $\{q_i^{(n)*}\}$.
	\end{algorithmic}
\end{algorithm}
  To deal with the bilinear terms in (\ref{cst_hatrho}), (\ref{cst_zeta}), and (\ref{cst_xi}), we utilize the relationship \( xy = \frac{1}{4} \left( (x+y)^2 - (x-y)^2 \right) \) to express each bilinear term as the difference of two convex terms and use Taylor expansion to get a convex approximation. In this way, (\ref{cst_hatrho}), (\ref{cst_zeta}), and (\ref{cst_xi}) are transformed as (\ref{cst_hatrho_sca}), (\ref{cst_zeta_sca}), and (\ref{cst_xi_sca}), respectively,
  \begin{figure*}[hbtp]
  \vspace{-0.6cm}
    \begin{align}
        \label{cst_hatrho_sca}
        &\frac{1}{4}\left((\hat{\rho}_i+g_i)^2-l_2(\hat{\rho}_i^{(n)},g_i^{(n)},\hat{\rho}_i,g_i)^2\right) -\frac{\sqrt{L}}{\log_2e}(\log_2(1+\zeta_i)-R_{c,i}) \le 0, i\in\mathcal{K},\\\vspace{-0.2cm}
        \label{cst_zeta_sca}
        & \sum_{j\ne i}
          \frac{1}{4} \left( (\zeta_i+q_j)^2 -l_2(\zeta_i^{(n)},q_j^{(n)},\zeta_i,q_j) \right) |\bar{\bm{h}}_{j}^H\bm{w}_{i}^u|^2 +
        \zeta_i-
        q_i|\bar{\bm{h}}_i^H\bm{w}_i^u|^2\le0,i\in\mathcal{K},\\\vspace{-0.2cm}
        \label{cst_xi_sca}
        & \sum_{j\ne i}
        \frac{1}{4} \left( (\xi_i-q_j)^2 - l_1(\xi_i^{(n)},q_j^{(n)},\xi_i,q_j) \right)|\bar{\bm{h}}_{j}^H\bm{w}_{i}^u|^2 -
        \xi_i+
        q_i|\bar{\bm{h}}_i^H\bm{w}_i^u|^2\le0,i\in\mathcal{K},\\
        \label{eqn_l1}
        & l_1(x^{(n)},y^{(n)},x,y) = 2(x^{(n)}+y^{(n)})(x-x^{(n)}+y-y^{(n)})+(x^{(n)}+y^{(n)})^2,\\
        \label{eqn_l2}
        & l_2(x^{(n)},y^{(n)},x,y) = 2(x^{(n)}-y^{(n)})(x-x^{(n)}-y+y^{(n)})+(x^{(n)}-y^{(n)})^2,
    \end{align}
    \vspace{-0.5cm}
    \hrulefill
    \vspace{-0.2cm}
    \end{figure*}
    where $\hat{\rho}_i^{(n)}$, $g_i^{(n)}$, $\zeta_i^{(n)}$, $\xi_i^{(n)}$,  and $q_i^{(n)}$ are the feasible solution from the $n$-th SCA iteration, $l_1$ and $l_2$  expressed as (\ref{eqn_l1}) and (\ref{eqn_l2})
    are the first order linear approximations for $(x+y)^2$ and $(x-y)^2$ at the operating point $(x^{(n)}, y^{(n)})$, respectively. 
Finally, for (\ref{cst_gi}), a convex upper bound for its LHS is derived using the first-order Taylor approximation to replace the concave terms, i.e., $\forall i\in\mathcal{K}$,
\begin{align}
\label{cst_gi_sca}
         &1-\frac{1}{(1+\xi_i)^2}-g_i^2\nonumber\\
         &\le 1+l_3(\xi_i^{(n)},\xi_i)-(g_i^{(n)2}+2g_i^{(n)}(g_i-g_i^{(n)})) ,
\end{align}
with 
$
    l_3(x^{(n)},x)\nonumber
    =\frac{2}{(1+x^{(n)})^3}(x-x^{(n)})-\frac{1}{(1+x^{(n)})^2}
$
being the first order Taylor approximation for $-\frac{1}{(1+x)^2}$ at the operating point $x^{(n)}$. 
Using the convex approximations to replace the LHSs of constraints (\ref{cst_t}), (\ref{cst_gi}), (\ref{cst_hatrho}), (\ref{cst_zeta}), and (\ref{cst_xi}), Problem {(P6.1)} is rewritten as
 \begin{align}
    \text{(P6.2)}
    \min_{\left\{\substack{
            q_i,t_i,\hat{\rho}_i,d_i,\\
            g_i,\zeta_i,\xi_i
        }\right\}}
    \quad
    & \sum_{i\in\mathcal{K}}\beta_i\tilde{d}_{k_i,s}(R_{s,i})+\frac{\beta_i\tilde{d}_{k_i,c}(R_{s,i})}
    {1+c_{k_i,i}t_{i}}\\
    \text{s.t.} \quad
    &U(t_i^{(n)},\hat{\rho}_i^{(n)},t_i,\hat{\rho}_i)\le0, i\in\mathcal{K},\\
    &\nonumber
    (\ref{cst_up_power}),
    (\ref{cst_bounds}),
    (\ref{cst_hatrho_sca}), 
    (\ref{cst_zeta_sca}),
    (\ref{cst_xi_sca}),
    (\ref{cst_gi_sca}),
\end{align}
Problem {(P6.2)} is a convex problem and can be easily solved using the classical convex optimization methods \cite{boyd2004convex}. Denote the local points at the $n$-th SCA iteration as $\bm{\Theta}^{(n)} = \{t_i^{(n)}, \hat{\rho}_i^{(n)}, d_i^{(n)},g_i^{(n)},\zeta_i^{(n)},\xi_i^{(n)}, q_i^{(n)}\}$, the SCA-based uplink power allocation algorithm is summarized in Algorithm \ref{alg_power}. The UDD-based algorithm for solving Problem {(P3)} is summarized in Algorithm \ref{alg_udd}.

\begin{algorithm}[hbt]
\caption{Joint Power and Beamforming Optimization Algorithm for Problem {(P3)}.}
\label{alg_udd}
\begin{algorithmic}[1]
\Require {$\{\bar{\bm{h}}_i,R_{s,i},R_{c,i},\beta_i\},P_{max}.$}
\Ensure {$\{p_i^*,\bm{w}_i^*\}$}
\State Set iteration number $n=1$.
\State Initialize the power and beamforming $\{p_i^{(n)},\bm{w}_i^{(n)}\}$ for Problem {(P4)}.
\State Convert the downlink power allocation $\{p_i^{(n)}\}$ to uplink power allocation $\{q_i^{(n)}\}$ using (\ref{eqn_p2q}) and set $\bm{w}_i^{u(n)}=\bm{w}_i^{(n)}$, $\forall i \in\mathcal{K}$.
\State \textbf{Repeat}
\State \quad Solve Problem {(P5)} for each user $i$ to obtain $\{\bm{w}_i^{u(n+1)}\}$ according to (\ref{eqn_mmse}) using $\{q_i^{(n)}\}$.
\State \quad Solve Problem {(P6)} to obtain $\{q_i^{(n+1)}\}$ according to Algorithm \ref{alg_power} using $\{\bm{w}_i^{u(n+1)}\}$.
        Update $n=n+1$.
\State \textbf{Until} \textnormal{the fractional decrease of the objective value of Problem {(P3)} is below a threshold $\epsilon_2$}.
\State Obtain $\{q_i^*,\bm{w}_i^{u*}\}$ as the convergent solution $\{q_i^{(n)},\bm{w}_i^{u(n)}\}$.
\State Convert $\{q_i^*\}$ to $\{p_i^*\}$ using (\ref{eqn_q2p}) and set $\bm{w}_i^{*}=\bm{w}_i^{u*}$, $\forall i \in\mathcal{K}$.\end{algorithmic}
\end{algorithm}

\subsection{Source Coding Rate Discretization}
Notice that alternatively solving Problems {(P2)} and {(P3)} until convergence does not solve Problem {(P1)} since the obtained source coding rates might not satisfy constraints (\ref{cst_data_rate}) and (\ref{cst_semantic_rate}). To address this, we use the round-down quantization strategy to obtain the discrete source coding rates. 
Denoting the convergent solution to Problems {(P2)} and {(P3)} as $\{\tilde{R}_{s,i}^*,\tilde{R}_{c,i}^*,\tilde{p}_i^*,\tilde{\bm{w}}_i^*\}$, we quantize $\tilde{R}_{s,i}^*$ to the nearest lower value in $\mathcal{R}_{k_i}$, i.e.,
$
{R}_{s,i}^{*}=\arg\max_{R\in\mathcal{R}_{k_i},R\le R_{s,i}^*}	R.
$
The channel coding rate is ${R}_{c,i}^{*}=\frac{{R}_{s,i}^{*}}{T_{i}}$. Using $\{{R}_{s,i}^{*},{R}_{c,i}^{*}\}$, we then compute the corresponding $\{p_i^{*},\bm{w}_i^{*}\}$ via Algorithm \ref{alg_udd}, initializing with $\{\tilde{p}_i^*,\tilde{\bm{w}}_i^*\}$ . The overall algorithm is summarized in Fig. \ref{fig_opt}.

\vspace{-0.3cm}   
\section{Experimental Results}
\label{section_experiment}
\begin{figure*}[htbp]
\centering
\subfloat[]{
\includegraphics[width=.34\linewidth]{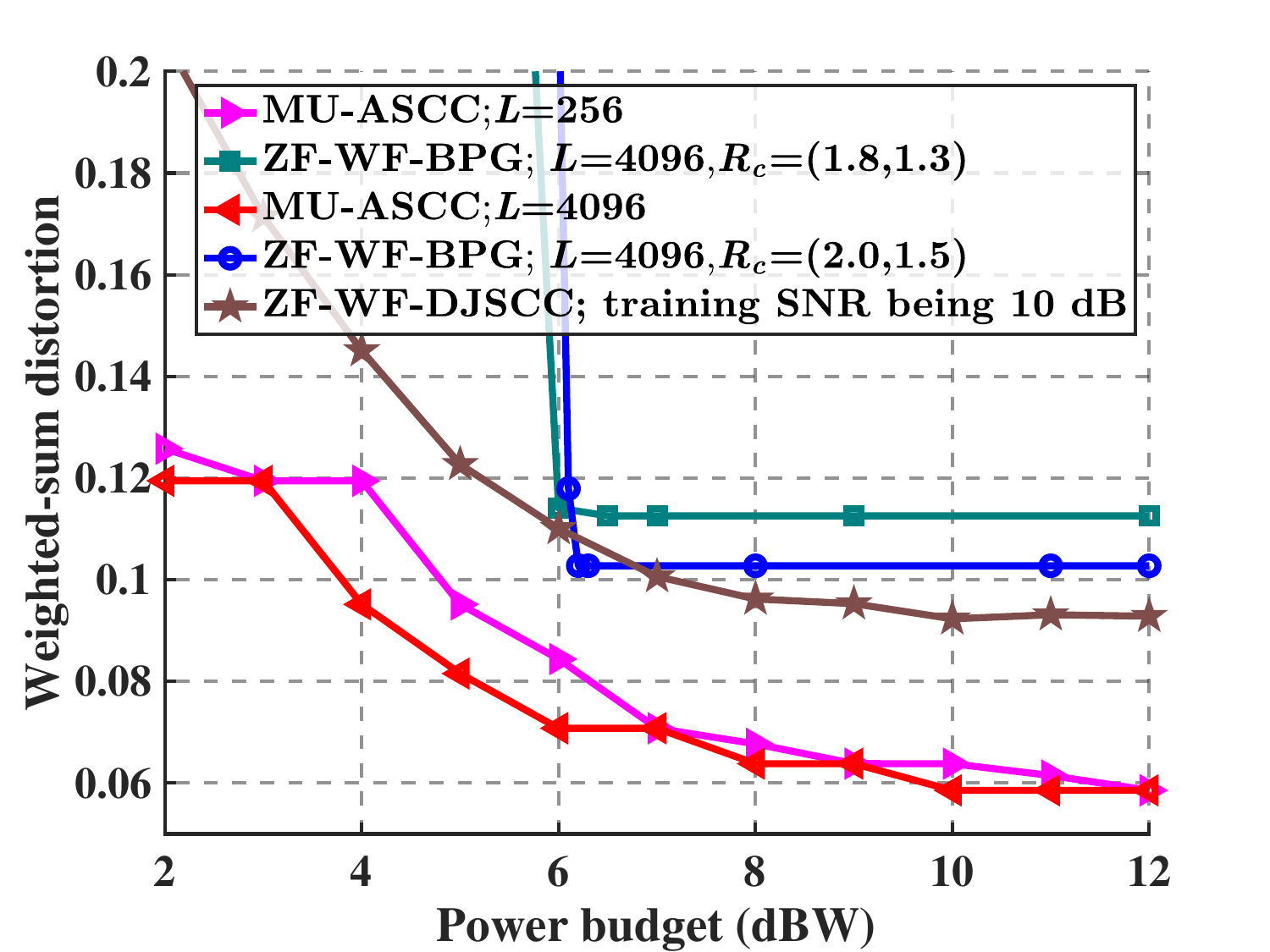}
\label{fig_wsdt_Pmax}
}\hspace{-0.8cm}
\subfloat[]{
\includegraphics[width=.34\linewidth]{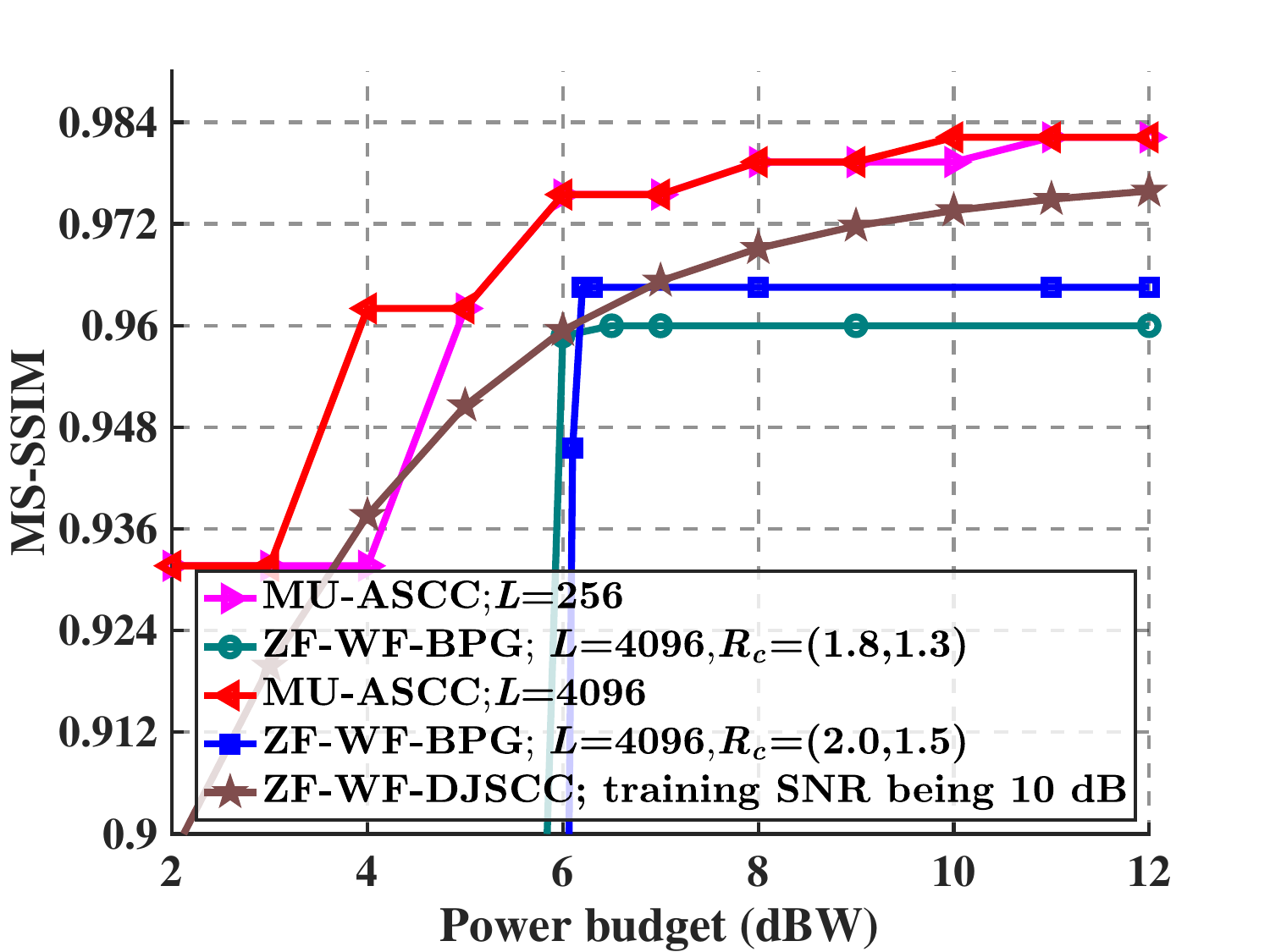}
\label{fig_ssim}
}\hspace{-0.8cm}
\subfloat[]{
\includegraphics[width=.34\linewidth]{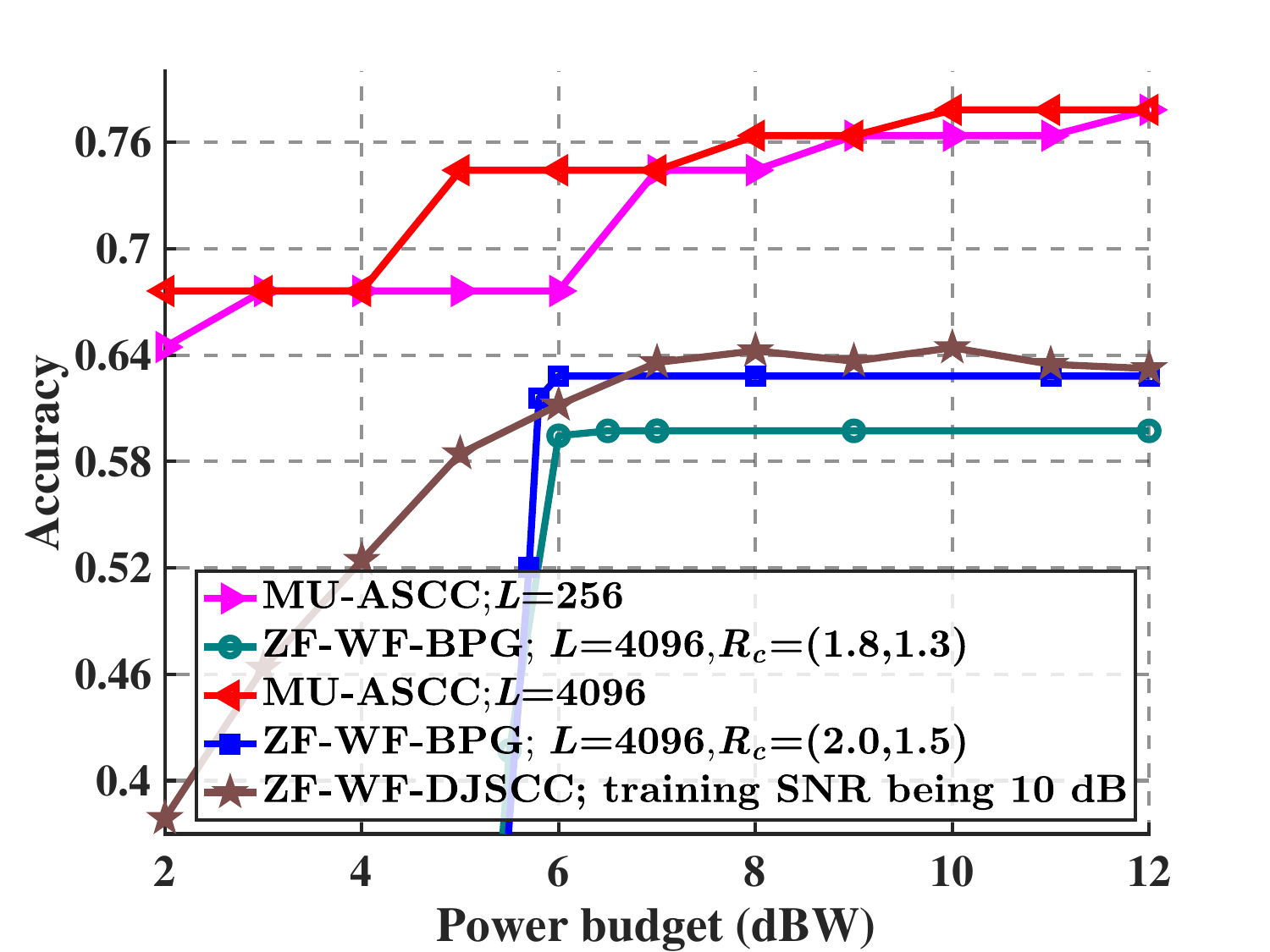}
\label{fig_acc}
}
\vspace{-0.2cm}
\caption{The comparison of DU and SU performances under different power budgets with blocklength being 256, 4096, respectively. The weights for the DU and SU are 0.8,0.2, respectively.}
\end{figure*}
\subsection{Experimental Settings}
\begin{itemize}
\item{\textbf{Datasets:}}
We consider the large-scale image dataset Caltech-UCSD Birds 200 (CUB-200-2011) to validate our proposed system. Specifically, the CUB-200-2011 dataset is a well-known dataset for bird photographs. It contains 11,788 images in 200 classes with image sizes up to 500$\times$500 pixels. 5,994 images are used for training and 5,794 images are used for testing. 
\item{\textbf{DNN Architecture and Hyperparameters:}}
We adopt the hyper-prior architecture \cite{balle2018variational} for our DNN design. The semantic encoder follows the inference model from \cite{balle2018variational}, and the DU decoder uses the corresponding synthesis model. We jointly train 23 image compression models covering source coding rates from $2.4\times 10^3$ to $9.3\times10^4$. For SUs, we construct the semantic decoder by fine-tuning a pre-trained ResNet-152 classifier \cite{he2016deep} together with the source decoder. 

Our framework is compatible with other image compression methods for reconstruction. For classification, semantic decoders can also extract labels directly from the bitstream without full image recovery. In this work, the simple concatenation-based decoder already yields significant performance gains as shown later.   

\item{\textbf{MU-MISO Channels:}}
We consider a two-user system consisting of a DU and an SU. The BS has 2 antennas. We consider the channel matrix as
\begin{align}
&\bar{\bm{H}} = [\bar{\bm{h}}_1,\bar{\bm{h}}_2] \nonumber \\
&\quad =\begin{bmatrix}
-0.4199 - 1.2885i, & -0.4546 + 1.0362i \\
0.2092 + 1.0851i, & -0.5603 + 0.7316i
\end{bmatrix}.\nonumber
\end{align}
Here we directly set the normalized channels $\bar{\bm{h}}_i$ for simplification and this is reasonable since any $\bm{h}_i$ and $\sigma_i$ setting can be easily converted to $\bar{\bm{h}}_i$. 

\item {\textbf{Benchmarking Schemes:}}
To validate the advantages of the proposed MU-ASCC scheme, we consider the following typical source and channel coding methods.
\begin{itemize}
	\item ZF-WF-BPG: This benchmark uses ZF beamforming and waterfilling (WF) power allocation \cite{cover1999elements}. Each user employs a fixed channel coding rate and BPG-based source coding. The semantic decoder concatenates the BPG decoder with the image classification network. 
	\item ZF-WF-DJSCC: This benchmark uses the ZF beamforming, WF power allocation with the DJSCC method \cite{bourtsoulatze2019deep} for image transmission. The semantic decoder combines the DJSCC decoder and the image classification network. We adopt this classical DJSCC method as the benchmark for fair comparison since its computational complexity is comparable to the adopted image compression technique \cite{balle2018variational}. Thus, the comparison focuses on source-channel adaptation and resource allocation rather than the advantages from newer semantic coding models.
\end{itemize}
\end{itemize}
\vspace{-0.37cm} 
\subsection{Multi-User Performance Trade-off}
\begin{figure}[h]
	\centering
	\includegraphics[width=0.8\linewidth]{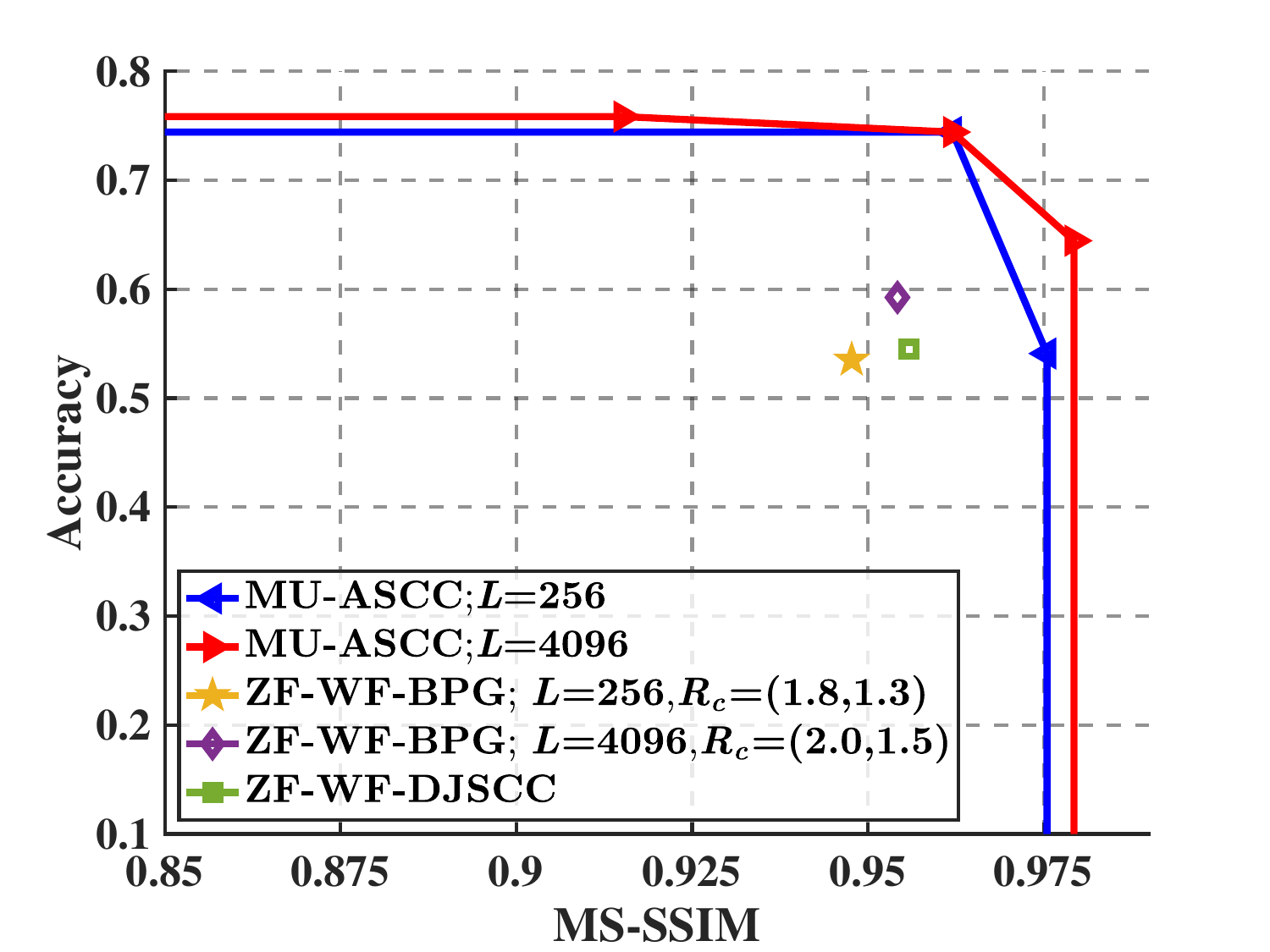}
	\vspace{-0.2cm}
	\caption{The achievable performance region for the MU-SemDaCom system along with the achievable performances of baselines.}
	\label{fig_performance_region}
\end{figure}
We evaluate the performance trade-off between two users using MS-SSIM and classification accuracy. Fig. \ref{fig_performance_region} shows the achievable performance region under different distortion weights, with a power budget of 3 Watts (W) and an average bandwidth ratio of 0.0356. The DJSCC model in the ``ZF-WF-DJSCC" scheme is trained and tested at the same signal-to-noise ratio (SNR) for each user. The proposed MU-ASCC scheme adaptively allocates resources by adjusting user weights, achieving a broader performance region than all benchmarks, indicating superior performance for both users simultaneously. This advantage stems from DNN-based semantic feature extraction and adaptive optimization of coding rates, power and beamforming. For example, at $L=256$, MU-ASCC improves classification by 21.20\% and MS-SSIM by 0.0143 compared to ``ZF-WF-BPG; $L$=256; $R_c$=(1.8,1.3)", and outperforms ``ZF-WF-DJSCC" by 17.23\% in accuracy and 0.0140 in MS-SSIM. Performance further improves with longer blocklength $L=4096$. 
\vspace{-0.3cm} 
\subsection{E2E Performance Comparisons}
\begin{figure*}[htbp]
\centering
\subfloat[]{
\includegraphics[width=.34\linewidth]{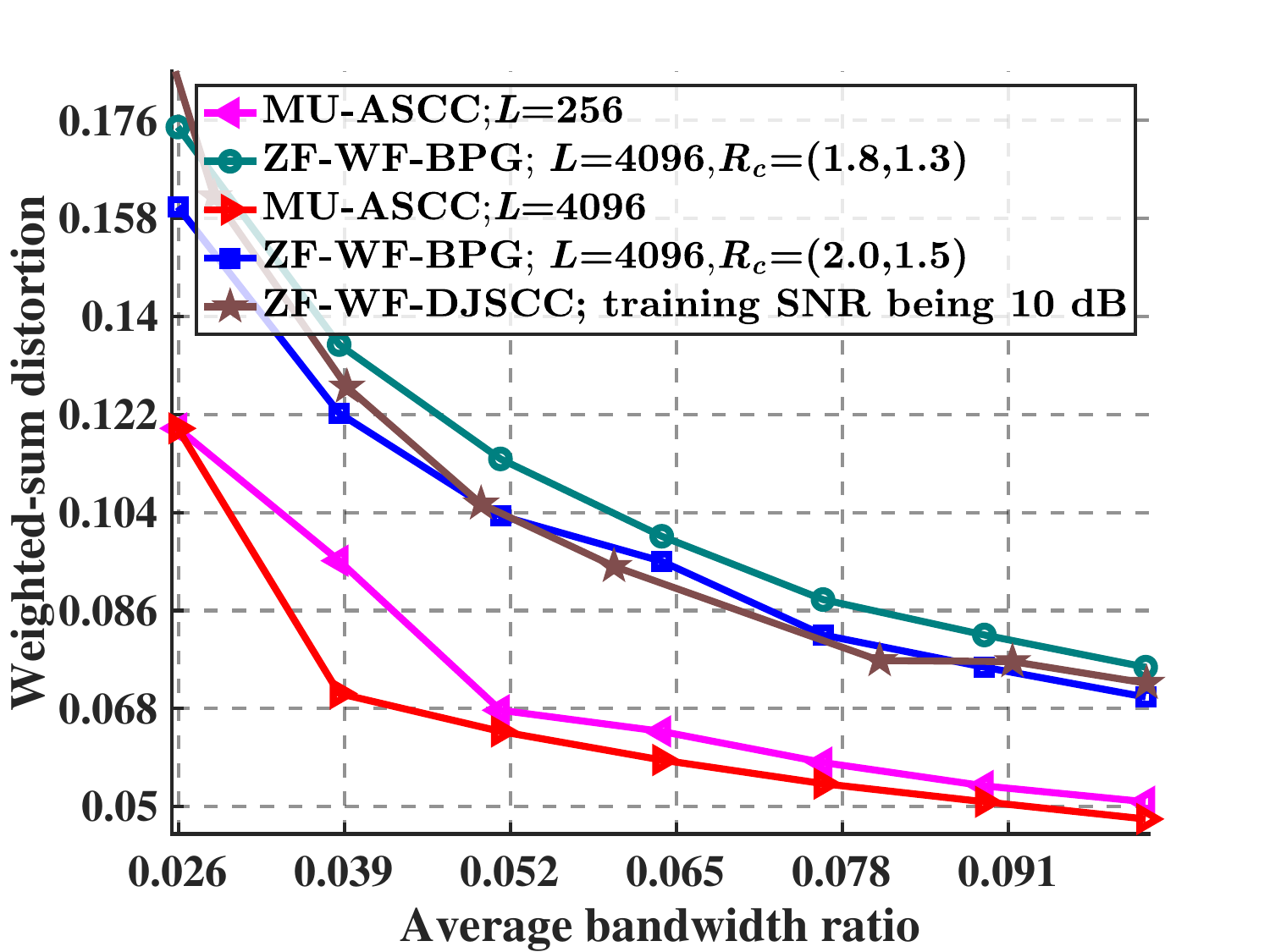}
\label{fig_delay_wsdt}
}\hspace{-0.9cm}
\subfloat[]{
\includegraphics[width=.34\linewidth]{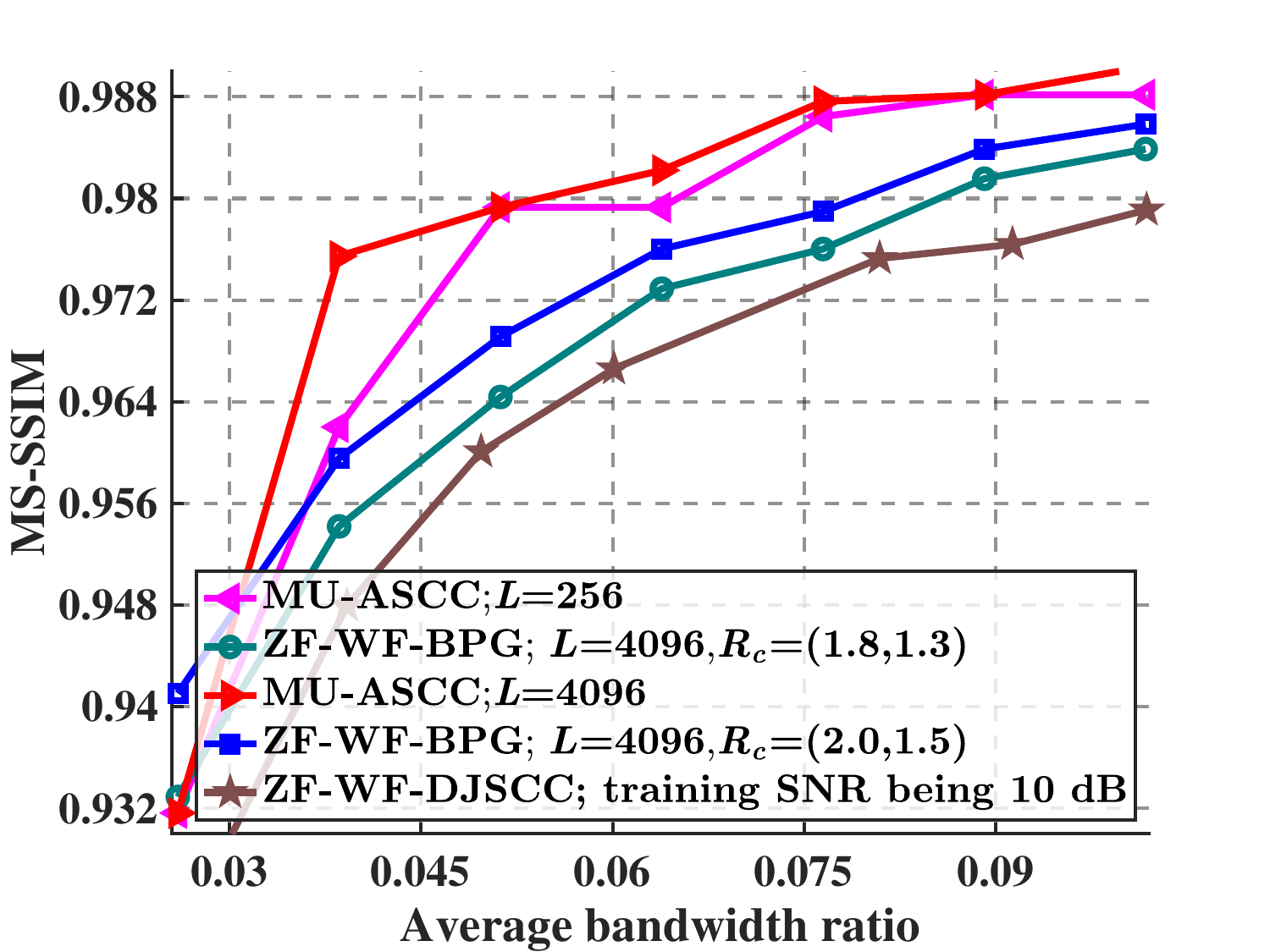}
\label{fig_delay_ssim}
}\hspace{-0.9cm}
\subfloat[]{
\includegraphics[width=.34\linewidth]{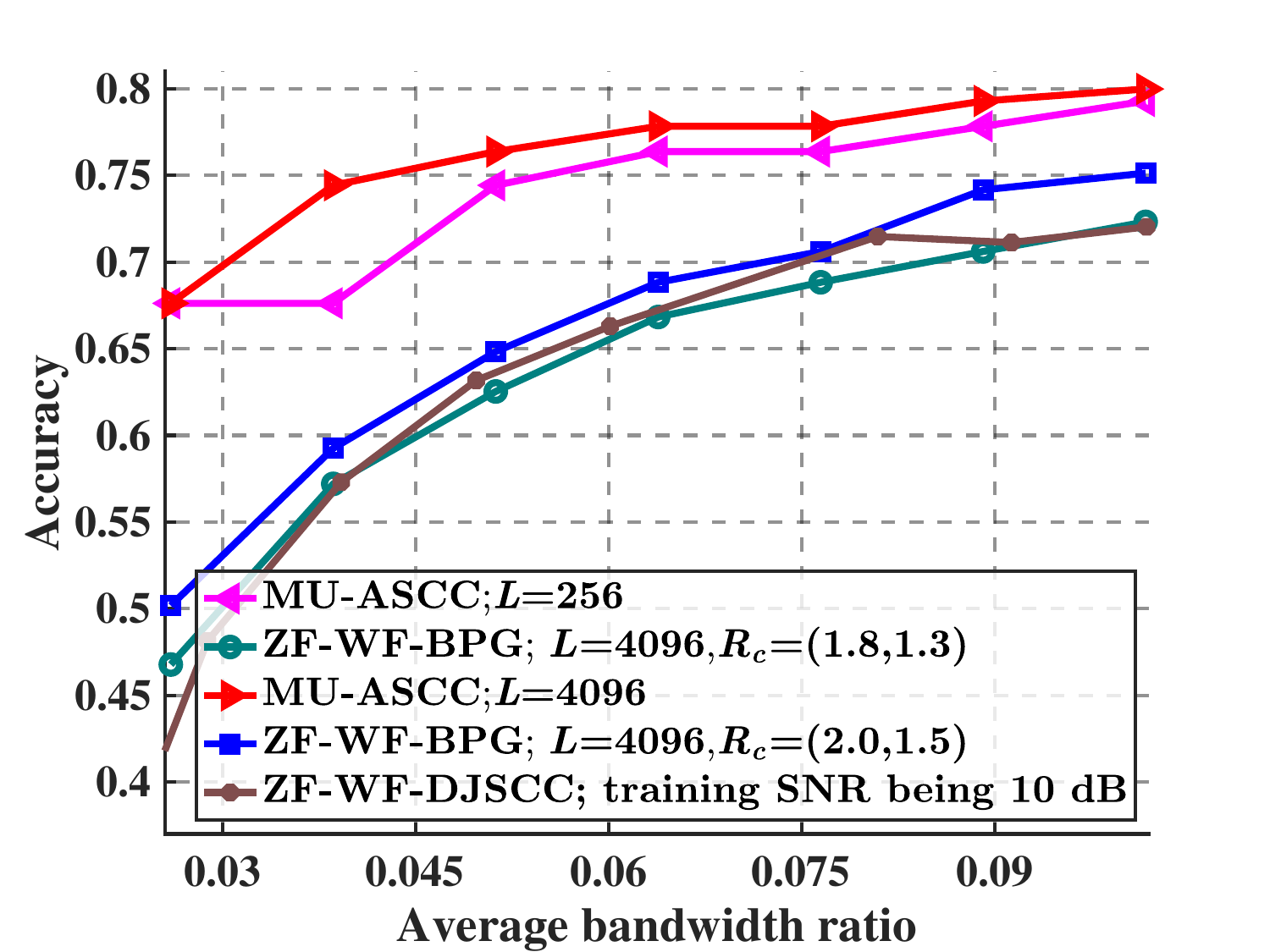}
\label{fig_delay_acc}
}
\vspace{-0.3cm}
\caption{The comparison of weighted-sum distortion with the bandwidth ratio. The weights for the DU and SU are $0.8,0.2$, respectively.}
\end{figure*}
We evaluate the E2E performance of the proposed method under varying power budgets and bandwidth ratios. 
For the ``ZF-WF-DJSCC" benchmark, as training separate DJSCC models for all possible channel conditions and every user is impractical in multi-user systems, we instead use a pre-trained DJSCC model (trained at 10 dB SNR, the typical SNR in our simulation settings) for this benchmark.

Fig. \ref{fig_wsdt_Pmax} shows the weighted-sum distortion versus power budget at an average bandwidth ratio of 0.0356 and user weights of 0.8 and 0.2. The proposed MU-ASCC scheme avoids the cliff effect and achieves significantly lower distortion than benchmarks by adaptively selecting coding rates to match the available power. For instance, at 8 dBW, MU-ASCC with $L=256$ outperforms ``ZF-WF-BPG; $L$=256; $R_c$=(1.8,1.3)" and ``ZF-WF-DJSCC with the training SNR being 10 dB" by 39.85\% and 29.63\%, respectively, while MU-ASCC with $L=4096$ surpasses corresponding benchmarks by 37.89\% and 33.67\%.

	Figs. \ref{fig_ssim} and \ref{fig_acc} present the MS-SSIM of the DU and classification accuracy of the SU versus power budget. MU-ASCC consistently outperforms benchmarks across nearly all power levels, owing to the strong representation ability of DNN-based codecs and the effective joint optimization of the JRPB algorithm.

	Finally, Fig. \ref{fig_delay_wsdt} illustrates weighted-sum distortion across bandwidth ratios under a 3 W power budget and user weights of 0.8 and 0.2. MU-ASCC maintains lower distortion than all benchmarks in all cases. Further, Fig. \ref{fig_delay_ssim} and Fig. \ref{fig_delay_acc} show that the proposed scheme achieves higher perceptual quality and task accuracy with less bandwidth. These gains stem from the expressive power of DNNs in feature extraction and semantic processing, as well as the ability of JRPB to adaptively allocate rates and powers under delay constraints and channel conditions while optimizing beamforming based on E2E distortion and user priorities.
\vspace{-0.3cm} 
\section{Concluding Remarks}
\label{section_conclusion}
This paper proposed an MU-ASCC scheme over MU-MISO channels, where the digital source and channel coding rates, power allocation and beamforming were jointly optimized to minimize the weighted-sum E2E distortion under the power budget and delay constraints. Specifically, we first proposed a MU-SemDaCom system over MU-MISO channels, which incorporates DUs aiming for data reconstruction and SUs focused on semantic task execution. Then, we built up the E2E distortion modeling for both the data recovery and the semantic task execution using the data regression method. Based on the MU-SemDaCom architecture and the E2E distortion modeling, we formulated an optimization problem to minimize the weighted-sum distortion by jointly optimizing the source and channel coding rates, power allocation, and beamforming. Finally, we proposed the JRPB algorithm to solve the optimization problem using the AO and SCA methods. Experimental results showed that the proposed MU-ASCC scheme outperformed the traditional DJSCC and SSCC schemes.
\vspace{-0.37cm} 
\appendices
%
  \vspace{-0.3 cm}
\section{Proof of Proposition \ref{prop_p3p4}}
\label{proof_prop_p3p4}
 To prove this proposition, we first show the following lemma:
 \begin{lemma}
 \label{lemma_ber_mono}
 The BER in (\ref{eqn_log10rho}) monotonically decreases as SINR $\gamma_i$ increases.
 \end{lemma}
\begin{IEEEproof}
	    Since the Q-function is monotonically decreasing, we analyze the monotonicity of the expression inside the Q-function (denoted as \(\hat{\rho}\)) with respect to \(\gamma_i\). The derivative of \(\hat{\rho}\) is
	\begin{equation}
		\frac{d\hat{\rho}}{d\gamma_i}=\frac{\sqrt{L}\,{\left(2\,\gamma_i -\log \left(\gamma_i +1\right)+R_{c,i} \,\log \left(2\right)+\gamma_i^2 \right)}}{{{\left(1-\frac{1}{{{\left(\gamma_i +1\right)}}^2 }\right)}}^{3/2} \,{{\left(\gamma_i +1\right)}}^3 }.
	\end{equation}
	Denote the numerator as $g$ and its derivative with respect to $\gamma_i$ is
	$
		\frac{dg}{d\gamma_i}=\sqrt{L}\,{\left(2\,\gamma_i -\frac{1}{\gamma_i +1}+2\right)}.
	$
	$\frac{dg}{d\gamma_i}$ monotonically increases and when $\gamma_i=0$, $\frac{dg}{d\gamma_i}=\sqrt{L}>0$. Therefore, $g>0$ and $\frac{d\hat{\rho}}{d\gamma_i}>0$ for $\gamma_i>0$, which implies $\hat{\rho}$ monotonically increases with $\gamma_i$. Given the Q-function decreases monotonically, it follows that $\tilde{\rho}_{b,i}$ in (\ref{eqn_log10rho}) decreases as $\gamma_i$ increases.
\end{IEEEproof}

    Now we prove Proposition \ref{prop_p3p4}. 
    We first show that Problems {(P3)} and {(P4)} have the same optimal value. For each feasible solution \(\{p_i,\bm{w}_i\}\) of {(P3)}, UDD theory transforms it to an uplink solution \(\{q_i,\bm{w}_i^u\}\) with \(\sum_{i=1}^K p_i = \sum_{i=1}^K q_i \leq P_{max}\). Thus, the optimal value \(D_4^*\) of Problem {(P4)} satisfies \(D_4^* \leq D_3^*\), where $D_3^*$ is the optimal value of Problem (P3). Similarly, any feasible solution of Problem {(P4)} corresponds to a feasible solution of Problem {(P3)}, implying \(D_3^* \leq D_4^*\). Therefore, \(D_3^* = D_4^*\).
    
    We now show that optimal solutions of Problems (P3) and (P4) are mutually transformable. Let \(\{p_i^*,\bm{w}_i^*\}\) be optimal for Problem (P3). By UDD theory, the transformed solution \(\{q_i^*,\bm{w}_i^{u*}\}\) satisfies \(\gamma_i^u = \gamma_i\), $i\in\mathcal{K}$. 
    Since Lemma \ref{lemma_ber_mono} and equations  (\ref{eqn_do_logis}) and (\ref{eqn_e2e_ds_rs_rho}) establish monotonic relationships between SINR and distortion, the objectives of Problems (P3) and (P4) coincide. Hence, \(\{q_i^*,\bm{w}_i^{u*}\}\) is optimal for Problem (P4). The converse holds similarly. 
      \vspace{-0.3 cm}
\section{Proof of Lemma \ref{lemma_sca_logQ}}
\label{proof_lemma_sca_logQ}
Since $\log(x)$ is concave, we have $\log(t_i)\le \frac{t_i}{t_i^{(n)}}+\log(t_i^{(n)})-1$ by its first order Taylor approximation at $t_i^{(n)}$. Similarly, to show $\log_{10}Q(\hat{\rho}_i)$ is upper bounded by its Taylor approximation $\tilde{Q}(\hat{\rho}_i^{(n)},\hat{\rho}_i)$, we prove $\hat{Q}(x)=\log Q(x)$ is concave for $x\ge 0$.
        The derivative of $\hat{Q}(x)$ is
	\begin{equation}
		\frac{d\hat{Q}}{dx} = \frac{-e^{-x^2/2}}{\int_x^\infty e^{-t^2/2}dt}
	\end{equation}
	and the second-order derivative is
	\begin{equation}
		\frac{d^2\hat{Q}}{dx^2} = 
		\frac{e^{-x^2/2}(x\int_x^\infty e^{-t^2/2}dt-e^{-x^2/2})}
		{(\int_x^\infty e^{-t^2/2}dt)^2}.
	\end{equation}
	Define 
	$
		f(x) = x\int_x^\infty e^{-t^2/2}dt-e^{-x^2/2}.
	$
	Then 
	$
		\frac{df}{dx} = \int_x^\infty e^{-t^2/2}dt > 0.
	$ When $x\rightarrow \infty$, using the Chernoff bound, 
	\begin{align}
		&f(x) \le x\sqrt{2\pi}e^{-x^2/2}-e^{-x^2/2},\\
		&=e^{-x^2/2}(\sqrt{2\pi}x-1)\rightarrow 0.
	\end{align}
	Hence $f\le 0$ for $x\ge 0$, implying $\frac{d^2\hat{Q}}{dx^2}\le 0$, so $\hat{Q}$ is concave. 
     Thus, both $\log(t_i)$ and $\log Q(\hat{\rho}_i)$ are bounded by their linear approximations, proving Lemma \ref{lemma_sca_logQ}.  
  \vspace{-0.3 cm}
\bibliographystyle{IEEEtran}
\bibliography{mu-semcom}
\end{document}